\documentclass[11pt,a4paper]{article}

\usepackage[utf8]{inputenc}
\usepackage[T1]{fontenc}
\usepackage{lmodern}
\usepackage[margin=1in]{geometry}

\usepackage{amsmath,amssymb}

\usepackage{booktabs}
\usepackage{array}

\usepackage{graphicx}
\usepackage{subcaption}
\usepackage{tikz}
\usepackage{pgfplots}
\pgfplotsset{compat=1.18}
\usepgfplotslibrary{fillbetween}
\usetikzlibrary{patterns,positioning}

\usepackage[colorlinks=true,linkcolor=blue,citecolor=blue,urlcolor=blue]{hyperref}

\usepackage[numbers,sort&compress]{natbib}

\usepackage{textcomp}

\usepackage{enumitem}

\title{Architecture-Aware LLM Inference Optimization on AMD Instinct GPUs:\\A Comprehensive Benchmark and Deployment Study}

\author{%
  Athos Georgiou\\
  \texttt{athos.georgiou@nca-it.com}
}

\date{Technical Report -- February 23, 2026}

\begin{document}

\maketitle

\begin{abstract}
Large language model (LLM) inference at frontier scale demands careful co-optimization of model architecture, hardware capabilities, and serving-system configuration, yet systematic benchmarking studies on AMD accelerators remain scarce.
We present a systematic cross-architecture evaluation of production LLM inference on AMD Instinct MI325X GPUs, benchmarking four models spanning 235 billion to 1 trillion parameters across three architectural families (MoE+MLA, Dense+GQA, MoE+GQA) on an 8-GPU cluster with 2\,TB aggregate HBM3e using vLLM~v0.14.1.
Our results demonstrate that architecture-aware optimization is essential. On the current ROCm stack, MLA models require block size~1 and cannot use KV cache offloading, while GQA models benefit from both. The AMD AITER runtime is required for competitive MLA inference throughput, with a Triton fallback available at substantially reduced performance, and must be selectively disabled for architectures with incompatible attention head configurations. A controlled AITER ablation on Llama-3.1-405B ($n{=}5$ per condition) reveals a modest 3--5\% throughput benefit at high concurrency but 2--16$\times$ higher measurement variability, confirming that AITER's large speedups target MoE/MLA kernels specifically.
Under text-only workloads, Llama-3.1-405B (Dense+GQA, 405B active) and DeepSeek~V3.2 (MoE+MLA, 37B active) achieve comparable peak throughput at 15,944 and 15,343 tok/s respectively, despite an order-of-magnitude difference in active parameters. Under vision workloads, Qwen3-VL-235B (MoE+GQA, 22B active) reaches 47,873 tok/s, 6.5$\times$ higher than Kimi-K2.5 (MoE+MLA, 32B active, 7,327 tok/s); these totals include image tokens from the vision encoder and are not directly comparable to the text-only results.
Active parameter count per token is associated with inference throughput across the models tested, though confounded by differences in quantization, AITER acceleration, and tensor parallelism (Section~\ref{sec:results}). All four models exhibit a common throughput saturation point within a given workload, consistent with a memory-bandwidth bottleneck (${\sim}$500 concurrent for short sequences, ${\sim}$100--200 for longer sequences).
All models maintain 100\% HTTP-level request success rates (HTTP~200 with valid response structure) through 1,000 concurrent users, processing 18.9 million tokens across 17{,}406 requests without failures.
\end{abstract}

\section{Introduction}
\label{sec:introduction}

Large language models (LLMs) have rapidly evolved from research prototypes into production infrastructure, powering applications ranging from conversational agents and code generation to scientific reasoning and multimodal understanding~\cite{vaswani2017attention, grattafiori2024llama3}. As model scale has grown from billions to trillions of parameters, the challenge of efficient inference serving has become a critical bottleneck for real-world deployment. While much attention has focused on training-side scaling laws, the operational reality of serving these models at scale, where throughput, latency, and hardware utilization must be simultaneously optimized, remains comparatively underexplored.

The inference landscape is further complicated by the rapid diversification of model architectures. Dense transformer models with Grouped-Query Attention (GQA)~\cite{ainslie2023gqa}, Mixture-of-Experts (MoE) architectures with hundreds of billions of total parameters~\cite{shazeer2017moe, jiang2024mixtral}, and novel attention mechanisms such as Multi-head Latent Attention (MLA)~\cite{deepseekv2_2024} each impose fundamentally different demands on memory hierarchy, parallelism strategies, and kernel-level optimization. A deployment configuration that yields excellent throughput for a dense GQA model may produce suboptimal or even erroneous results when applied to an MoE model with MLA, yet systematic studies of these architecture-specific serving behaviors remain scarce.

Simultaneously, the GPU accelerator landscape is broadening beyond a single vendor. AMD's Instinct MI325X accelerators, based on the CDNA~3 architecture, offer 256~GB of HBM3e memory per device with 6.0~TB/s bandwidth, presenting a compelling alternative for large-scale inference workloads. The ROCm software ecosystem has matured to support production inference frameworks such as vLLM~\cite{kwon2023vllm}, yet comprehensive benchmarking studies on AMD hardware for state-of-the-art LLMs, particularly trillion-parameter models, remain limited in scope.

In this work, we present a cross-architecture benchmark study of LLM inference on AMD Instinct MI325X GPUs, systematically evaluating four architecturally diverse models spanning from 235~billion to 1~trillion parameters using the vLLM serving framework. Our testbed comprises an 8-GPU cluster with 2~TB of aggregate HBM3e, and our evaluation covers concurrency scaling from single-request latency through 1{,}000 concurrent users, stress testing under diverse workload profiles, and architecture-specific optimization. The models under study comprise Kimi-K2.5 (1T total, 32B active; MoE+MLA)~\cite{kimik2_2025}, DeepSeek~V3.2 (685B total, 37B active; MoE+MLA)~\cite{deepseekv3_2024, deepseekv32_2025}, Llama-3.1-405B (405B dense; GQA)~\cite{grattafiori2024llama3}, and Qwen3-VL-235B (235B total, 22B active; MoE+GQA)~\cite{qwen3_2025}, collectively representing the major architectural paradigms in contemporary LLM design.

Our study makes the following key contributions:

\begin{enumerate}
    \item \textbf{Cross-architecture MI325X benchmark study for LLM inference at scale.} We provide the first academic cross-architecture evaluation of production LLM serving on AMD Instinct MI325X hardware with architecture-specific configuration characterization, covering four frontier models across three architectural families (dense GQA, MoE+GQA, MoE+MLA), with workloads ranging from single requests to 1{,}000 concurrent users. Industry benchmarks such as SemiAnalysis InferenceMAX~\cite{semianalysis2025inferencemax} provide multi-model MI325X throughput comparisons, but do not characterize the architecture-specific configuration constraints that govern deployment. All configurations achieve 100\% success rates under stress testing, processing 18.9~million tokens across 17{,}406 requests without failures.

    \item \textbf{Architecture-aware optimization requires fundamentally different strategies.} We demonstrate that MoE, MLA, and GQA architectures demand distinct serving configurations. On the current ROCm stack, MLA models require block size~1 and are incompatible with KV cache offloading, while GQA models benefit from KV offloading and standard block sizes. The AMD AI Tensor Engine for ROCm (AITER) is required for competitive production MLA inference throughput on ROCm; a Triton MLA fallback exists but delivers substantially lower performance, making AITER a practical necessity for production deployments. A controlled A/B ablation on the GQA-based Llama-3.1-405B ($n{=}5$ independent server restarts per condition) reveals a modest 3--5\% throughput benefit at high concurrency but a 2--16$\times$ increase in measurement variability (coefficient of variation), confirming that AITER's documented 2--3$\times$ speedups are specific to MoE and MLA kernels rather than general attention acceleration. AITER must be disabled for MLA configurations with incompatible head-count constraints. These findings challenge the assumption that a single serving configuration can be applied across architectures.

    \item \textbf{Trillion-parameter model deployment on production GPU clusters.} We demonstrate successful deployment and benchmarking of Kimi-K2.5, a 1~trillion parameter MoE model, on a cluster of 4~MI325X GPUs using INT4 quantization-aware training (QAT) weights. This represents, to our knowledge, the first published inference benchmark of a trillion-parameter model on MI325X (CDNA~3), achieving 7{,}327~tok/s throughput at 500 concurrent requests with 100\% reliability.

    \item \textbf{Empirical validation that active parameters drive throughput at frontier scale.} It is well established in MoE literature that active parameter count per token, not total parameter count, governs inference compute cost~\cite{jiang2024mixtral, chittyvenkata2025moeinferencebench}. Our results provide quantitative validation of this principle at frontier scale on MI325X, though the comparison is confounded by other experimental variables (Section~\ref{sec:results}). Qwen3-VL-235B (22B active) achieves 47{,}873~tok/s despite having more total parameters than models with lower throughput, while the 1T-parameter Kimi-K2.5 (32B active) achieves throughput comparable to models one-third its total size. This empirical confirmation at frontier scale has practical implications for model selection in throughput-sensitive deployments.

    \item \textbf{Workload-dependent throughput saturation.} All four models, despite spanning a 4$\times$ range in total parameters and employing three different architectural paradigms, exhibit a common throughput saturation point within a given workload on our 8-GPU MI325X cluster. The saturation threshold is workload-dependent: ${\sim}$500 concurrent for the stress test workload (500-token input, 200-token output), confirmed by fine-grained sweeps from 500 to 1{,}000 in steps of 50, and ${\sim}$100--200 concurrent for longer-sequence workloads (2{,}048-token input, 512-token output). This common saturation across architecturally diverse models within a given workload suggests a memory-bandwidth bottleneck, consistent with the DRAM bandwidth saturation dynamics described by Recasens et al.~\cite{arXiv_2503_08311}.
\end{enumerate}

The remainder of this paper is organized as follows. Section~\ref{sec:related_work} surveys related work in LLM serving systems, model architectures, and quantization techniques. Section~\ref{sec:architecture} describes the system architecture and hardware platform. Section~\ref{sec:optimization} details the optimization techniques employed. Section~\ref{sec:methodology} presents our experimental methodology, including workload design and metrics. Section~\ref{sec:results} reports our experimental results across all models and workload configurations. Section~\ref{sec:discussion} discusses the implications of our findings, and Section~\ref{sec:conclusion} concludes with directions for future work.

\section{Related Work}
\label{sec:related_work}

Our work sits at the intersection of LLM serving systems, model architecture innovation, hardware-aware optimization, and quantization for inference. We survey each area and position our contributions within the existing literature.

\subsection{LLM Serving Systems}

The efficient serving of large language models has emerged as a critical systems challenge, driven by the autoregressive nature of text generation and the enormous memory footprint of modern models.

\paragraph{Continuous batching and iteration-level scheduling.}
Orca~\cite{yu2022orca} introduced iteration-level scheduling for transformer-based generative models, enabling the serving system to add and remove requests from a batch at each decoding iteration rather than waiting for an entire batch to complete. This continuous batching approach, combined with selective batching that applies batching only to compatible operations, achieved up to 36.9$\times$ throughput improvement over NVIDIA FasterTransformer on GPT-3 175B. Orca's design established the foundational scheduling paradigm adopted by subsequent serving systems.

\paragraph{Memory-efficient KV cache management.}
Building on Orca's scheduling innovations, vLLM~\cite{kwon2023vllm} addressed the critical memory management challenge for KV caches. The PagedAttention algorithm, inspired by virtual memory and paging in operating systems, partitions the KV cache of each sequence into fixed-size blocks that can be stored in non-contiguous memory. This approach achieves near-zero memory waste (under 4\%), enabling 2--4$\times$ throughput improvements over prior systems including Orca and FasterTransformer. vLLM has become the de facto standard for LLM serving, supporting a wide range of models and hardware backends including AMD ROCm. Our work uses vLLM as the serving framework and provides a systematic cross-architecture evaluation of its performance on MI325X hardware across architecturally diverse frontier models, complementing industry benchmarks from SemiAnalysis InferenceMAX~\cite{semianalysis2025inferencemax} and MLPerf~\cite{reddi2020mlperf} with architecture-specific configuration characterization.

\paragraph{Alternative serving frameworks.}
SGLang~\cite{zheng2024sglang} introduced RadixAttention for KV cache reuse across structured language model programs, achieving up to 5$\times$ throughput over vLLM for workloads with prefix sharing. HuggingFace Text Generation Inference (TGI) provides a production-ready serving solution with continuous batching and tensor parallelism support. NVIDIA's TensorRT-LLM offers vendor-specific optimizations including custom attention kernels, inflight batching, and FP8/FP4 quantization for NVIDIA GPUs. While these systems have been extensively benchmarked on NVIDIA hardware, systematic evaluations on AMD accelerators remain limited. Our study fills this gap by providing detailed performance characterization of vLLM on MI325X across multiple model architectures.

\subsection{Model Architectures}

The models evaluated in this study span three major architectural paradigms: dense transformers with grouped-query attention, mixture-of-experts models, and models employing multi-head latent attention. Each architecture presents distinct serving challenges.

\paragraph{The transformer foundation.}
The transformer architecture~\cite{vaswani2017attention} established self-attention as the dominant paradigm for sequence modeling. As models scaled to hundreds of billions of parameters, the KV cache required for autoregressive generation became a primary memory bottleneck, motivating architectural innovations to reduce its footprint.

\paragraph{Attention mechanism variants.}
Multi-Query Attention (MQA)~\cite{shazeer2019mqattention} proposed sharing key and value heads across all query heads, dramatically reducing KV cache size and memory bandwidth requirements during decoding. Grouped-Query Attention (GQA)~\cite{ainslie2023gqa} generalized this approach by using an intermediate number of key-value heads, achieving quality close to standard multi-head attention with speed approaching MQA. GQA has been widely adopted in production models including the Llama family~\cite{grattafiori2024llama3}. In our evaluation, Llama-3.1-405B employs GQA in a dense architecture, while Qwen3-VL-235B combines GQA with MoE. We find that GQA models benefit from KV cache offloading and standard PagedAttention block sizes, achieving the highest absolute throughput in our study.

\paragraph{Multi-head Latent Attention (MLA).}
DeepSeek-V2~\cite{deepseekv2_2024} introduced Multi-head Latent Attention, which compresses the KV cache into a low-rank latent representation, reducing KV cache memory by 93.3\% compared to standard multi-head attention while maintaining or improving model quality. MLA has been adopted by DeepSeek-V3~\cite{deepseekv3_2024} and Kimi-K2.5~\cite{kimik2_2025}. However, MLA introduces significant serving constraints: on the current ROCm stack it requires block size~1 for PagedAttention and is incompatible with KV cache offloading, and imposes specific attention head distribution requirements for tensor parallelism and hardware-accelerated kernels. While individual constraints have been documented in scattered GitHub issues (vLLM, AITER), our work consolidates these into a systematic characterization across multiple MLA models on MI325X and demonstrates their cumulative impact on deployment configuration.

\paragraph{Mixture-of-Experts (MoE).}
The Sparsely-Gated Mixture-of-Experts architecture~\cite{shazeer2017moe} demonstrated that model capacity can scale independently of computational cost by routing each token to a sparse subset of expert networks. Mixtral~8x7B~\cite{jiang2024mixtral} popularized MoE for LLMs, using 8~experts with 2~selected per token to achieve 47B total parameters with only 13B active. DeepSeek-V3~\cite{deepseekv3_2024} (subsequently updated as V3.2~\cite{deepseekv32_2025}) scaled this to 685B parameters with 37B active, while Kimi-K2.5~\cite{kimik2_2025} reaches 1~trillion parameters with 384~experts and 32B~active per token. It is well established that active parameter count, rather than total count, governs per-token inference compute~\cite{jiang2024mixtral, chittyvenkata2025moeinferencebench}. Our benchmarks corroborate this at frontier scale on MI325X across the models tested, though the comparison is confounded by other experimental variables (see Section~\ref{sec:results} for details), with practical implications for model selection in production deployments.

\subsection{Hardware Acceleration for LLM Inference}

\paragraph{Attention kernel optimization.}
FlashAttention~\cite{dao2022flashattention} introduced an IO-aware exact attention algorithm that tiles attention computation to minimize HBM reads and writes, achieving 2--4$\times$ wall-clock speedups for transformer training and inference. FlashAttention-2~\cite{dao2023flashattention2} further improved parallelism and work partitioning, reaching 50--73\% of theoretical peak FLOPs on NVIDIA A100 GPUs. These kernel-level optimizations are critical for inference throughput, and their availability (or absence) on specific hardware platforms directly impacts serving performance.

\paragraph{AMD Instinct and ROCm ecosystem.}
AMD's Instinct MI300X and MI325X accelerators, based on the CDNA~3 architecture, provide high-bandwidth memory (up to 256~GB HBM3e at 6.0~TB/s per device on MI325X) designed for large-scale AI workloads. The ROCm (Radeon Open Compute) software stack provides an open-source platform for GPU computing, including support for PyTorch, vLLM, and custom kernel libraries. The AMD AI Tensor Engine for ROCm (AITER) provides optimized kernels for MoE and attention operations on Instinct hardware, with AMD reporting 2--3$\times$ inference speedups for compatible architectures~\cite{amd2024aiter}. However, AITER compatibility is architecture-dependent. On the current ROCm stack, AITER is required for competitive production MLA inference throughput, consistent with AMD's documentation and community experience; a Triton MLA fallback exists but delivers substantially lower performance, making controlled ablation impractical. We find AITER must be entirely disabled for Kimi-K2.5 due to MXFP4 hardware requirements (CDNA~4 only) and attention head count constraints (the AITER MLA backend supports exactly 16 or 128 heads per rank; supported values may vary by AITER version). Prior benchmarking studies of LLM inference on AMD hardware have been limited in scope, typically evaluating single models or narrow workload ranges. Our work provides the first academic multi-model evaluation spanning dense, MoE, and MLA architectures on MI325X with architecture-specific configuration characterization.

\subsection{Quantization for LLM Inference}

Model quantization has become essential for deploying large models within GPU memory constraints, with techniques spanning post-training quantization (PTQ) and quantization-aware training (QAT).

\paragraph{Weight quantization.}
GPTQ~\cite{frantar2023gptq} demonstrated accurate post-training quantization to 3--4 bits per weight using approximate second-order information, enabling 175B-parameter models to fit in a single GPU for the first time. AWQ~\cite{lin2024awq} introduced activation-aware weight quantization, identifying that protecting only 1\% of salient weight channels (determined by activation distributions) significantly reduces quantization error. AWQ received the MLSys 2024 Best Paper Award and has become widely adopted for INT4 weight-only quantization.

\paragraph{Weight-activation quantization.}
SmoothQuant~\cite{xiao2023smoothquant} enabled W8A8 (8-bit weights and activations) quantization by smoothing activation outliers through an offline mathematical transformation that migrates quantization difficulty from activations to weights. FP8 quantization has emerged as a practical format for LLM inference, with recent studies demonstrating that FP8 weight and activation quantization (W8A8-FP) is effectively lossless across model scales for the Llama family. FP8 quantization reduces memory consumption by approximately 50\% compared to FP16/BF16, enabling larger models and batch sizes on fixed hardware.

\paragraph{Quantization in our study.}
Our evaluation spans multiple quantization strategies dictated by architectural constraints: FP8 for Llama-3.1-405B and DeepSeek~V3.2, BF16 for Qwen3-VL-235B (whose vision encoder dimensions are incompatible with FP8 block quantization kernels on both ROCm and CUDA platforms), and INT4 QAT compressed tensors for Kimi-K2.5. We find that quantization format selection is not merely a precision-performance trade-off but is constrained by architecture-specific compatibility with hardware kernels: an underappreciated dimension of deployment planning.

\subsection{KV Cache Optimization}

Efficient KV cache management is central to high-throughput LLM serving. Beyond PagedAttention's memory-efficient allocation~\cite{kwon2023vllm}, recent work has explored KV cache compression, eviction, and offloading strategies~\cite{yuan2024kvcache_survey}. KV cache offloading to CPU memory extends effective context capacity for memory-bound workloads, and vLLM supports this through its native offloading backend. However, we demonstrate that KV cache offloading compatibility is architecture-dependent: GQA models (Llama-3.1-405B, Qwen3-VL-235B) successfully utilize offloading, while MLA models (DeepSeek~V3.2, Kimi-K2.5) are incompatible with the offloading connector on the current ROCm stack. vLLM's planned offloading redesign (RFC~\#22605) would also benefit MLA models by supporting their compressed latent KV cache format. This architectural dependency has significant implications for context length scaling and workload planning.

\subsection{Positioning of This Work}

The MLPerf Inference benchmark~\cite{reddi2020mlperf} provides a standardized framework for cross-platform ML inference evaluation. AMD submitted MI325X results for Llama~2~70B in MLPerf Inference v5.0, with v5.1 (September 2025) expanding coverage to include Mixtral~8x7B and Llama~2~70B Interactive scenarios. However, the benchmark suite does not currently cover the full architectural diversity examined in our study: specifically, MoE routing, multi-head latent attention, and trillion-parameter model deployment.

LLM-Inference-Bench~\cite{chittyvenkata2024llminferencebench} benchmarks eight models across seven hardware platforms, including AMD MI300X, using vLLM and TensorRT-LLM, but evaluates only models up to 72B parameters, does not include MI325X, and does not reach 1{,}000-user concurrency. SemiAnalysis InferenceMAX~\cite{semianalysis2025inferencemax} provides nightly multi-model, multi-hardware throughput-latency Pareto frontiers including MI325X, but does not characterize architecture-specific configuration constraints or provide fixed-concurrency scaling analysis. MoE-Inference-Bench~\cite{chittyvenkata2025moeinferencebench} systematically evaluates MoE inference throughput across varying active/total parameter ratios, but covers only models up to 70B on NVIDIA H100. Existing academic LLM serving benchmarks have otherwise predominantly focused on NVIDIA hardware with single-architecture evaluations. Our work contributes the first academic cross-architecture comparative benchmark on MI325X at 1{,}000-concurrent-user scale, encompassing (1)~four architecturally diverse frontier models (dense GQA, MoE+GQA, MoE+MLA), (2)~characterization of architecture-specific constraints governing serving configuration (block size, AITER compatibility, KV offloading, tensor parallelism), (3)~the first published inference benchmark of a trillion-parameter model on MI325X (CDNA~3), acknowledging AMD's prior MI355X benchmarks and ORNL's MI250X training work, and (4)~empirical characterization of workload-dependent throughput saturation behavior (consistent with the memory-bandwidth bottlenecks modeled analytically by LIMINAL~\cite{davies2025liminal} and observed empirically by Recasens et al.~\cite{arXiv_2503_08311}) across all four models on our 8-GPU MI325X cluster. Speculative decoding~\cite{leviathan2023speculativedecoding}, an orthogonal optimization for autoregressive generation, is not evaluated in this study but represents a promising direction for future work.

\section{System Architecture and Hardware Platform}
\label{sec:architecture}

This section describes the hardware platform, inference engine, and model architectures evaluated in our study. Our experimental setup targets a production-representative configuration comprising an 8-GPU AMD Instinct MI325X cluster running vLLM v0.14.1, serving four frontier-scale large language models spanning diverse architectural families.

\subsection{AMD Instinct MI325X Platform}
\label{subsec:mi325x}

All experiments were conducted on a single-node server equipped with eight AMD Instinct MI325X accelerators. The MI325X is built on AMD's CDNA~3 architecture (gfx942, the instruction set architecture identifier for CDNA~3) and represents the current generation of AMD data center GPUs optimized for large-scale AI inference workloads. Table~\ref{tab:mi325x-specs} summarizes the key specifications.

\begin{table}[ht]
\centering
\caption{AMD Instinct MI325X specifications and 8-GPU cluster aggregate resources.}
\label{tab:mi325x-specs}
\begin{tabular}{lcc}
\toprule
\textbf{Specification} & \textbf{Per GPU} & \textbf{8-GPU Cluster} \\
\midrule
HBM3e Capacity & 256\,GB & 2\,TB \\
Memory Bandwidth & 6.0\,TB/s & 48\,TB/s \\
FP16 Compute & 1,307\,TFLOPS & 10.5\,PFLOPS \\
Architecture & \multicolumn{2}{c}{CDNA~3 (gfx942)} \\
\bottomrule
\end{tabular}
\end{table}

The MI325X's 256\,GB HBM3e capacity per accelerator is a critical enabler for frontier-scale model serving. With tensor parallelism across eight GPUs, even the largest models leave substantial per-GPU headroom: DeepSeek~V3.2 in FP8 requires only ${\sim}$83\,GiB of weight memory per GPU (as reported by vLLM's model loading log), consuming approximately 35\% of each GPU's 256\,GB capacity (noting that 83\,GiB $\approx$ 89\,GB) and leaving the remainder available for KV cache and batch state. This eliminates the need for KV cache offloading to CPU memory in most deployment scenarios, reducing architectural complexity and latency.

LLM inference is fundamentally memory-bandwidth-bound rather than compute-bound~\cite{pope2023efficiently}. The MI325X's 6.0\,TB/s memory bandwidth per accelerator (48\,TB/s aggregate) directly translates to higher token throughput, particularly at large batch sizes where memory access patterns dominate runtime. The CDNA~3 microarchitecture provides hardware-level support for FP8 matrix operations, enabling quantized inference without dedicated quantization accelerators.

The system runs ROCm~6.4.2 with RCCL~2.26.6 for multi-GPU communication. NUMA balancing is disabled (\texttt{kernel.numa\_balancing=0}) to prevent the operating system from migrating memory pages between NUMA nodes, which would degrade GPU communication latency. The software stack is containerized using Docker~29.1.5 with ROCm support, ensuring reproducible deployments.

\subsection{vLLM Inference Engine}
\label{subsec:vllm}

We use vLLM~v0.14.1~\cite{kwon2023vllm} as the inference serving engine. vLLM implements several key techniques that enable efficient high-throughput LLM serving:

\paragraph{PagedAttention.} vLLM's core memory management innovation is PagedAttention, which manages KV cache memory using a paging mechanism inspired by virtual memory systems in operating systems. Rather than pre-allocating contiguous memory blocks for each sequence's KV cache, PagedAttention divides the cache into fixed-size blocks that can be allocated and freed independently. This virtually eliminates memory fragmentation and enables near-optimal memory utilization, allowing more concurrent sequences to be served from the same GPU memory budget.

\paragraph{Continuous Batching.} Unlike static batching approaches that wait for an entire batch to complete before processing new requests, vLLM implements continuous (or iteration-level) batching. New requests are inserted into the running batch at each decoding step as slots become available, maximizing GPU utilization and minimizing queuing delays.

\paragraph{V1 Engine and Chunked Prefill.} vLLM~v0.14.1 uses the V1 engine architecture, in which chunked prefill is always enabled and cannot be disabled. Chunked prefill splits long prompt processing into smaller chunks that are interleaved with decode steps from other sequences. This prevents long prompts from monopolizing the GPU and allows decode-phase sequences to maintain low inter-token latency even when new long-context requests arrive. The chunk size is controlled by \texttt{--max-num-batched-tokens}, which determines the maximum number of tokens processed in a single scheduler iteration.

\paragraph{Multi-Step Scheduling.} vLLM supports multi-step scheduling via the \texttt{--num-scheduler-\allowbreak{}steps} parameter, which batches multiple decode steps before returning to the scheduler. This reduces CPU-side scheduling overhead and improves GPU utilization, with values of 10--15 providing measurable throughput gains before diminishing returns.

All models are served through vLLM's OpenAI-compatible API endpoint, containerized in Docker images (\texttt{vllm/vllm-openai-\allowbreak{}rocm:latest} for stable models; \texttt{rocm/vllm-dev:\allowbreak{}nightly} for Kimi-K2.5 which requires a nightly build, specifically v0.16.0rc1.dev88).

\subsection{Model Architectures}
\label{subsec:models}

We evaluate four frontier-scale models spanning three architectural families: Mixture-of-Experts with Multi-head Latent Attention (MoE+MLA), dense transformer with Grouped-Query Attention (Dense+GQA), and MoE with GQA (MoE+GQA). Table~\ref{tab:model-specs} provides a comparative overview.

\begin{table*}[ht]
\centering
\caption{Evaluated model architectures. Active parameters are per-token for MoE models. Per-GPU memory is from vLLM startup logs on the 8-GPU MI325X cluster at the deployed precision and tensor parallelism.}
\label{tab:model-specs}
\footnotesize
\setlength{\tabcolsep}{5pt}
\begin{tabular}{lccccccc}
\toprule
\textbf{Model} & \textbf{Total} & \textbf{Active} & \textbf{Architecture} & \textbf{Attention} & \textbf{Context} & \textbf{Precision} & \textbf{Per-GPU} \\
 & \textbf{Params} & \textbf{Params} & & & \textbf{Length} & & \textbf{Memory} \\
\midrule
DeepSeek V3.2 & 685B & $\sim$37B & MoE + MLA & MLA & 160K & FP8 & $\sim$83\,GiB \\
Llama-3.1-405B & 405B & 405B & Dense + GQA & GQA & 128K & FP8 & $\sim$112\,GiB \\
Qwen3-VL-235B & 235B & $\sim$22B & MoE + GQA & GQA & 256K & BF16 & $\sim$58\,GiB \\
Kimi-K2.5 & 1T & $\sim$32B & MoE + MLA & MLA & 256K & INT4 QAT & $\sim$145\,GiB \\
\bottomrule
\end{tabular}
\end{table*}

\paragraph{DeepSeek~V3.2 (685B).} DeepSeek~V3.2~\cite{deepseekv3_2024, deepseekv32_2025} is a Mixture-of-Experts model that uses Multi-head Latent Attention (MLA) to compress key-value pairs into a low-rank latent space, reducing the per-token KV cache footprint relative to standard multi-head attention. The model activates approximately 37B of its 685B total parameters per token via its expert routing mechanism. On our MI325X cluster, DeepSeek~V3.2 requires mandatory configuration of \texttt{--block-size~1} for the MLA KV cache format and the \texttt{AITER\_ENABLE\_VSKIP=0} environment variable to prevent \texttt{HSA\_STATUS\_ERROR\_MEMORY\_APERTURE\_VIOLATION} errors in fused MoE kernels on CDNA~3. The model is served in FP8 precision with AITER acceleration enabled.

\paragraph{Llama-3.1-405B.} Meta's Llama-3.1-405B-Instruct~\cite{grattafiori2024llama3} is the largest dense (non-MoE) transformer in our evaluation. It employs Grouped-Query Attention (GQA), which shares key-value heads across multiple query heads to reduce KV cache memory requirements while maintaining attention quality. As a dense model, all 405B parameters are activated for every token, making it the most compute-intensive model per token in our study. FP8 quantization is essential to fit this model within the 8-GPU cluster's memory budget while leaving room for KV cache at practical context lengths.

\paragraph{Qwen3-VL-235B.} Qwen3-VL-235B-A22B-Instruct~\cite{qwen3_2025} is a vision-language model combining a Mixture-of-Experts language backbone with a vision encoder (ViT). The MoE architecture activates only $\sim$22B of its 235B total parameters per token, yielding the lowest active parameter count in our evaluation. The model uses GQA for its attention mechanism. A notable constraint is that its vision encoder's intermediate MLP dimension of 4304 is not divisible by 128 (the block size used by FP8 block quantization kernels), making it incompatible with FP8 quantization. This is a model-architecture constraint, not a platform-specific limitation; the same incompatibility has been reported on NVIDIA hardware (vLLM Issues~\#30934, \#26589). Consequently, it must be served in BF16 precision. Despite this, its low active parameter count enables the highest throughput among all evaluated models.

\paragraph{Kimi-K2.5 (1T).} Moonshot AI's Kimi-K2.5~\cite{kimik2_2025} is the largest model in our study at 1~trillion total parameters. It employs an MoE architecture with 384 experts (8 selected per token), activating approximately 32B parameters per forward pass. Like DeepSeek~V3.2, it uses MLA for attention. However, AITER cannot be enabled for Kimi-K2.5 on MI325X: enabling AITER encounters a fatal error during CUDA graph capture (``MXFP4 is not available on your device''), because the AITER MLA backend's MXFP4 (microscaling FP4) quantization pathway requires hardware support available only on MI350X and later (CDNA~4). Additionally, Kimi-K2.5's 64-head MLA configuration creates a tensor parallelism constraint: with TP$=$8, each GPU receives only 8 attention heads, which falls outside the AITER MLA backend's supported head counts (exactly 16 or 128 per rank; supported values may vary by AITER version). The official Moonshot AI deployment guide specifies TP$=$8 with the Triton MLA fallback at reduced performance. In our evaluation, Kimi-K2.5 is deployed with TP$=$4 and AITER disabled, utilizing only half of the available GPUs; the TP$=$4 choice was made to match the AITER head count constraint (64~/ 4 = 16 heads per GPU) during initial deployment attempts before the MXFP4 limitation was identified, and was retained for consistency. The model ships with native INT4 Quantization-Aware Training (QAT) via the \texttt{compressed-tensors} format, including a 400M-parameter MoonViT vision encoder. It requires the vLLM nightly build (\texttt{rocm/vllm-dev:nightly}) as stable releases do not yet support its architecture.

Table~\ref{tab:model-config} summarizes the deployment configuration for each model, including the AITER component flags, tensor parallelism degree, and block size required by the architecture.

\begin{table}[ht]
\centering
\caption{Per-model deployment configuration on the MI325X cluster. AITER columns indicate which kernel components are enabled (1) or disabled (0).}
\label{tab:model-config}
\begin{tabular}{lcccccc}
\toprule
\textbf{Model} & \textbf{AITER} & \textbf{MHA} & \textbf{MLA} & \textbf{MoE} & \textbf{Block} & \textbf{TP} \\
 & & & & & \textbf{Size} & \\
\midrule
DeepSeek V3.2 & 1 & 0 & 1 & 1 & 1 & 8 \\
Llama-3.1-405B & 1\textsuperscript{a} & 1 & 0 & 0 & 16 & 8 \\
Qwen3-VL-235B & 1\textsuperscript{a} & 1 & 0 & 1 & 16 & 8 \\
Kimi-K2.5 & 0\textsuperscript{b} & -- & -- & -- & 1 & 4 \\
\bottomrule
\end{tabular}

\smallskip
\noindent\textsuperscript{a}\,For Llama and Qwen3-VL, AITER is enabled explicitly (\texttt{VLLM\_ROCM\_USE\_AITER=1}). Note that \texttt{VLLM\_ROCM\_USE\_AITER} defaults to \texttt{0} (disabled) when unset. DeepSeek also sets it explicitly.\\
\noindent\textsuperscript{b}\,Kimi-K2.5 requires AITER disabled (\texttt{VLLM\_ROCM\_USE\_AITER=0}) due to attention head count incompatibility; AITER=1 encounters a fatal error during CUDA graph capture.
\end{table}

Detailed command-line flags for each model's deployment are provided in Table~\ref{tab:model-flags} (Section~\ref{sec:methodology}).

The architectural diversity across these four models, spanning dense and sparse parameter activation, MLA and GQA attention mechanisms, text-only and vision-language modalities, and FP8, BF16, and INT4 precision formats, provides a comprehensive testbed for evaluating the AMD MI325X platform's versatility as an inference accelerator for frontier-scale models.

\section{Optimization Techniques}
\label{sec:optimization}

Deploying frontier-scale models on the MI325X cluster requires a combination of quantization, memory management, kernel-level acceleration, parallelism configuration, and scheduling optimizations. This section details the techniques applied and their model-specific considerations.

\subsection{Quantization Strategies}
\label{subsec:quantization}

Quantization reduces the numerical precision of model weights and activations, decreasing memory consumption and improving computational throughput at the cost of potential accuracy degradation. We employ three distinct quantization strategies depending on model architecture and compatibility constraints, summarized in Table~\ref{tab:quantization}.

\begin{table}[ht]
\centering
\caption{Quantization strategy and compatibility for each model. FP8 KV refers to whether \texttt{--kv-cache-dtype fp8} can be used.}
\label{tab:quantization}
\begin{tabular}{lcccc}
\toprule
\textbf{Model} & \textbf{Method} & \textbf{Precision} & \textbf{FP8 KV} & \textbf{Memory Savings} \\
\midrule
DeepSeek V3.2 & FP8 & W8A8 & No\textsuperscript{a} & $\sim$50\% \\
Llama-3.1-405B & FP8 & W8A8 & Yes\textsuperscript{b} & $\sim$50\% \\
Qwen3-VL-235B & None (BF16) & W16A16 & No\textsuperscript{c} & -- \\
Kimi-K2.5 & INT4 QAT & W4 & No\textsuperscript{d} & $\sim$75\% \\
\bottomrule
\end{tabular}

\raggedright
\footnotesize
\textsuperscript{a}MLA backend uses \texttt{fp8\_ds\_mla} format automatically; \texttt{--kv-cache-dtype fp8} is incompatible.\\
\textsuperscript{b}FP8 KV cache is architecturally supported but was not enabled in our benchmarks; savings reflect FP8 weights only.\\
\textsuperscript{c}Vision encoder intermediate dimension (4304) not divisible by 128 (FP8 block quantization block size); incompatible with FP8 kernels on both ROCm and CUDA platforms.\\
\textsuperscript{d}MLA architecture incompatible with standard FP8 KV cache format.
\end{table}

\paragraph{FP8 Quantization (W8A8).} Standard FP8 quantization (\texttt{--quantization fp8}) converts both weights and activations from 16-bit to 8-bit floating point, yielding approximately 50\% memory reduction with minimal accuracy loss for most workloads~\cite{micikevicius2022fp8}. On the AMD ROCm platform, we additionally evaluated Per-Token Per-Channel FP8 (PTPC-FP8, \texttt{--quantization ptpc\_fp8}), which applies per-token scaling to activations and per-channel scaling to weights. PTPC-FP8 provides improved accuracy over standard FP8 by adapting the dynamic range to each token's activation distribution and each output channel's weight distribution independently, and is the recommended FP8 mode for ROCm since vLLM~v0.7.3.

The MI325X's CDNA~3 architecture implements the OCP FP8 standard, using the E4M3 format (4 exponent bits, 3 mantissa bits, dynamic range $\pm$448) for both weights and activations during inference~\cite{micikevicius2022fp8}. Two FP8 quantization modes are available in vLLM for ROCm: per-tensor scaling (\texttt{fp8}) and per-token-per-channel scaling (\texttt{ptpc\_fp8}). Both achieve identical memory reduction; PTPC-FP8 provides better numerical accuracy by adapting dynamic range independently per token and output channel. Our benchmarks use per-tensor \texttt{fp8} quantization.

FP8 is critical for fitting Llama-3.1-405B on the cluster: at BF16 the per-GPU weight memory would be approximately 224\,GiB (extrapolating from the measured FP8 footprint), leaving minimal headroom within each GPU's 256\,GB HBM for KV cache and runtime buffers. With FP8, the per-GPU weight footprint drops to ${\sim}$112\,GiB, enabling deployment with substantial KV cache capacity. For DeepSeek~V3.2, FP8 reduces the per-GPU weight footprint from an estimated ${\sim}$180\,GiB to ${\sim}$83\,GiB.

\paragraph{INT4 Quantization-Aware Training.} Kimi-K2.5 ships with native INT4 QAT quantization using the \texttt{compressed-tensors} format~\cite{kimik2_2025}. Unlike post-training quantization, QAT incorporates quantization effects during training, enabling more aggressive compression (4-bit weights) while preserving model quality. This reduces the 1T-parameter model to ${\sim}$145\,GiB per GPU (TP$=$4), fitting comfortably within four MI325X GPUs (1\,TB aggregate HBM at TP$=$4). No additional quantization flags are needed; vLLM automatically detects the compressed-tensors format.

\paragraph{Vision Encoder Constraints.} Qwen3-VL-235B's vision encoder (ViT) contains MLP layers with dimensions not divisible by 128 (e.g., the intermediate dimension of 4304, where $4304 / 128 = 33.625$), which is the block size requirement for FP8 block quantization kernels. This is a model-architecture constraint rather than a platform-specific limitation: the same dimension incompatibility has been reported on NVIDIA hardware (vLLM Issues~\#30934, \#26589). Attempting FP8 quantization produces a runtime error due to this alignment constraint. The model must therefore be served entirely in BF16. Despite this precision penalty, the MoE architecture's low active parameter count ($\sim$22B) enables competitive throughput.

\subsection{KV Cache Management}
\label{subsec:kv-cache}

The KV cache stores key and value tensors from prior tokens during autoregressive generation and often dominates GPU memory consumption at long context lengths and high concurrency. We evaluate two complementary strategies: KV cache quantization and KV cache offloading to CPU memory.

\paragraph{FP8 KV Cache.} For models with GQA-based attention, the KV cache can be independently quantized to FP8 (\texttt{--kv-cache-dtype fp8}), reducing its memory footprint by approximately 50\% relative to BF16. This is orthogonal to weight quantization and can be combined with FP8 weights for cumulative savings. Our Llama-3.1-405B benchmarks used FP8 weight quantization only; FP8 KV cache was not enabled, yielding approximately 50\% weight memory savings.

However, FP8 KV cache is incompatible with both MLA-based models in our evaluation. DeepSeek~V3.2's \texttt{ROCMAiterMLASparseBackend} uses a specialized \texttt{fp8\_ds\_mla} format that is automatically selected by vLLM; specifying \texttt{--kv-cache-dtype fp8} produces a \texttt{ValueError}. Similarly, Kimi-K2.5's MLA implementation does not support the standard FP8 KV cache interface.

\paragraph{CPU Offloading.} vLLM supports offloading KV cache blocks to CPU memory via \texttt{--kv-offloading-\allowbreak{}backend native}, with a configurable buffer size (\texttt{--kv-offloading-size}). This extends effective KV cache capacity beyond GPU HBM at the cost of increased latency from CPU--GPU data transfers. In our experiments, we configure offloading buffers of 64\,GiB for models that support it.

On the current ROCm stack (vLLM~v0.14.1), KV cache offloading is not supported for MLA models, producing a \texttt{KeyError} at runtime. GQA models successfully use offloading on the same ROCm stack, indicating this limitation is MLA-specific rather than a blanket ROCm constraint. vLLM's planned offloading redesign (RFC~\#22605), which proposes a separated-process architecture with CUDA IPC handles, would also benefit MLA models by supporting their compressed latent KV cache format. Table~\ref{tab:kv-offload} summarizes compatibility.

\begin{table}[ht]
\centering
\caption{KV cache optimization compatibility by model and attention architecture.}
\label{tab:kv-offload}
\begin{tabular}{lcccc}
\toprule
\textbf{Model} & \textbf{Attention} & \textbf{FP8 KV} & \textbf{CPU Offload} & \textbf{Block Size} \\
\midrule
DeepSeek V3.2 & MLA & No & No & 1 \\
Llama-3.1-405B & GQA & Yes & Yes & 16 \\
Qwen3-VL-235B & GQA & No\textsuperscript{a} & Yes & 16 \\
Kimi-K2.5 & MLA & No & No & 1 \\
\bottomrule
\end{tabular}

\raggedright
\footnotesize
\textsuperscript{a}FP8 KV is architecturally compatible with GQA but blocked by the vision encoder's FP8 incompatibility.
\end{table}

In practice, the MI325X's 256\,GB HBM capacity per GPU (2\,TB aggregate) substantially reduces the need for KV cache offloading. For the models and context lengths evaluated (up to 32K tokens with 1{,}000 concurrent requests), all workloads fit within HBM without offloading. We note, however, that offloading flags also serve as a workaround for a GEMM kernel compatibility error (\texttt{RuntimeError:\ wrong!\ device\_gemm \ldots\ does not support this GEMM problem}) that affects Llama-3.1-405B and Qwen3-VL-235B under certain AITER configurations.

\subsection{AITER Kernel Optimization}
\label{subsec:aiter}

AMD's AI Tensor Engine for ROCm (AITER) provides optimized compute kernels specifically designed for the Instinct GPU family. AITER replaces generic GPU kernels with hand-tuned implementations that exploit CDNA~3 microarchitectural features, yielding substantial performance improvements for key inference primitives.

\paragraph{Performance Impact.} According to AMD's documentation, AITER provides the following speedups for DeepSeek V3/R1 workloads~\cite{amd2024aiter}: 2.1$\times$ overall inference acceleration, 2$\times$ faster block-scale GEMM operations, and 3$\times$ faster fused Mixture-of-Experts execution. These gains are most pronounced for MoE models where the expert routing and sparse matrix operations are dominant computational bottlenecks. We note that on the current ROCm stack, AITER is required for competitive production MLA inference throughput; a Triton MLA fallback exists (\texttt{VLLM\_ROCM\_USE\_AITER=0}) but delivers substantially lower performance, making AITER a practical necessity for production deployments. See Section~\ref{subsec:optimization-impact} for details.

\paragraph{Component-Level Control.} AITER is activated via the master environment variable \texttt{VLLM\_ROCM\_\allowbreak{}USE\_AITER=1} (which defaults to \texttt{0}/disabled when unset). When enabled, it activates the following component flags:

\begin{itemize}
\item \texttt{VLLM\_ROCM\_USE\_AITER\_LINEAR} -- Quantization and GEMM operations
\item \texttt{VLLM\_ROCM\_USE\_AITER\_MOE} -- Fused Mixture-of-Experts kernels
\item \texttt{VLLM\_ROCM\_USE\_AITER\_RMSNORM} -- Accelerated RMS normalization
\item \texttt{VLLM\_ROCM\_USE\_AITER\_MHA} -- Multi-Head Attention kernels
\item \texttt{VLLM\_ROCM\_USE\_AITER\_MLA} -- Multi-head Latent Attention kernels
\item \texttt{VLLM\_ROCM\_USE\_AITER\_FP8BMM} -- FP8 batched matrix multiplication
\end{itemize}

Individual components can be selectively disabled when encountering model-specific incompatibilities (e.g., \texttt{VLLM\_ROCM\_USE\_AITER\_MLA=0} for MLA accuracy issues under certain TP/DP configurations).

\paragraph{VSKIP Configuration Requirement.} A known issue (AITER Issue~\#1143, October 2025) affects AITER on MI300X/MI325X hardware: the \texttt{AITER\_ENABLE\_\allowbreak{}VSKIP} flag defaults to \texttt{true} when unset, causing \texttt{HSA\_STATUS\_ERROR\_MEMORY\_APERTURE\_VIOLATION} errors during inference with DeepSeek models. This must be explicitly disabled (\texttt{AITER\_ENABLE\_VSKIP=0}) for stable operation. The root cause is an incompatibility in the VSKIP-enabled fused MoE kernel variant on CDNA~3 hardware. We confirm this workaround is required on MI325X with vLLM~v0.14.1.

\paragraph{Model-Specific AITER Behavior.} AITER's applicability varies significantly across models (see Table~\ref{tab:model-config}). DeepSeek~V3.2 benefits from full AITER acceleration (MLA + MoE kernels) with the VSKIP workaround. Llama-3.1-405B uses AITER's MHA kernels but not MLA or MoE, as it is a dense GQA model. Qwen3-VL-235B uses AITER's MHA and MoE kernels. Kimi-K2.5, however, must run with AITER entirely disabled (\texttt{VLLM\_ROCM\_\allowbreak{}USE\_AITER=0}). Enabling AITER encounters a fatal error during CUDA graph capture (``MXFP4 is not available on your device''): the AITER MLA backend's MXFP4 (microscaling FP4) quantization pathway requires hardware support available only on MI350X and later (CDNA~4), not on MI325X (CDNA~3). Independently, its 64-head MLA configuration also falls outside the AITER MLA backend's supported head counts (exactly 16 or 128 per rank; supported values may vary by AITER version).

\paragraph{FP8 BMM Warmup.} AITER's FP8 batched matrix multiplication kernels require pre-compilation on first invocation, adding approximately 3 minutes of warmup time to the initial model load for FP8-quantized models. Subsequent starts use cached compiled kernels and incur no additional overhead.

\subsection{Tensor Parallelism Configuration}
\label{subsec:tensor-parallelism}

Tensor parallelism (TP)~\cite{shoeybi2019megatronlm} distributes model layers across multiple GPUs by partitioning weight matrices along specific dimensions, enabling models that exceed single-GPU memory to be served with low latency. All models in our evaluation require multi-GPU serving, but the optimal TP configuration differs based on architectural constraints.

\paragraph{Default TP$=$8 Configuration.} Three of the four models (DeepSeek~V3.2, Llama-3.1-405B, and Qwen3-VL-235B) are deployed with TP$=$8, utilizing all eight MI325X GPUs. This maximizes the available memory bandwidth (48\,TB/s aggregate) and memory capacity, and is the natural configuration for the 8-GPU cluster. Inter-GPU communication uses RCCL~2.26.6 with NCCL minimum channels set to 112 (\texttt{NCCL\_MIN\_\allowbreak{}NCHANNELS=112}) for high-throughput all-reduce operations. For large TP configurations, quantized all-reduce (\texttt{VLLM\_ROCM\_QUICK\_\allowbreak{}REDUCE\_QUANTIZATION=FP}) and BF16-to-FP16 casting (\texttt{VLLM\_ROCM\_QUICK\_\allowbreak{}REDUCE\_CAST\_BF16\_TO\_FP16=1}) provide additional communication bandwidth savings.

\paragraph{Kimi-K2.5 TP$=$4 Constraint.} Kimi-K2.5 represents a notable exception to the TP$=$8 default. The model's MLA architecture has 64 attention heads. Under TP$=$8, each GPU would receive $64 / 8 = 8$ heads, which falls outside the AITER MLA backend's supported head counts (exactly 16 or 128 per rank; supported values may vary by AITER version). With AITER disabled, TP$=$8 is possible using the Triton MLA fallback (as specified in the official Moonshot AI deployment guide), but at reduced performance. In our evaluation, we use TP$=$4 with AITER disabled. The TP$=$4 choice was made during initial deployment to match the AITER head count constraint ($64 / 4 = 16$ heads per GPU) before the MXFP4 hardware limitation was identified (Section~\ref{subsec:aiter}), and was retained for consistency across benchmark runs. We note that TP$=$8 with the Triton MLA fallback would be a valid alternative that utilizes the full cluster's bandwidth. Using only four GPUs, Kimi-K2.5 has access to only 1\,TB of HBM and 24\,TB/s of aggregate bandwidth, compared to 2\,TB and 48\,TB/s for TP$=$8 models.

\paragraph{Block Size.} MLA-based models (DeepSeek~V3.2 and Kimi-K2.5) require \texttt{--block-size~1} for their KV cache allocation on the ROCm/AITER stack. This is a platform-specific constraint: on NVIDIA hardware, FlashMLA uses block size~64, and Intel Gaudi uses block size~128 for MLA models. GQA-based models (Llama-3.1-405B and Qwen3-VL-235B) use the default block size of 16, which amortizes allocation overhead and enables more efficient memory management.

\subsection{Concurrency and Scheduling}
\label{subsec:concurrency}

Maximizing throughput under concurrent load requires tuning vLLM's batching and scheduling parameters. The key parameters are \texttt{--max-num-seqs} (maximum concurrent sequences per batch), \texttt{--max-num-\allowbreak{}batched-tokens} (maximum tokens per scheduler iteration), and \texttt{--num-scheduler-\allowbreak{}steps} (decode steps per scheduling round).

\paragraph{Batch Size Configuration.} The \texttt{--max-num-seqs} parameter controls the upper bound on concurrent sequences. We evaluate configurations ranging from low-latency (128--256 sequences) to maximum-throughput (2048--4096 sequences). For our high-throughput configuration, we set \texttt{--max-num-seqs~2048} with \texttt{--max-num-batched-\allowbreak{}tokens~65536}, which provides sufficient batch capacity to saturate the MI325X cluster's memory bandwidth. Table~\ref{tab:scheduling} summarizes the configuration profiles.

\begin{table}[ht]
\centering
\caption{Scheduling configuration profiles and their trade-offs.}
\label{tab:scheduling}
\begin{tabular}{lccc}
\toprule
\textbf{Profile} & \textbf{max-num-seqs} & \textbf{max-batched-tokens} & \textbf{scheduler-steps} \\
\midrule
Low latency & 256 & 8,192 & 1 \\
Balanced & 512 & 16,384 & 10 \\
High throughput & 1,024 & 32,768 & 15 \\
Maximum & 2,048 & 65,536 & 15 \\
\bottomrule
\end{tabular}
\end{table}

\paragraph{Multi-Step Scheduling.} The \texttt{--num-scheduler-steps} parameter batches multiple decode iterations before the scheduler re-evaluates the batch composition. Setting this to 15 reduces CPU-side scheduling overhead, as the scheduler runs once per 15 decode steps rather than once per step. Values above 20 yield diminishing returns, as the overhead savings plateau while responsiveness to new requests decreases.

\paragraph{GPU Memory Utilization.} The \texttt{--gpu-memory-utilization} parameter controls the fraction of GPU HBM allocated to the KV cache pool after model weights are loaded. We use 0.95 for maximum-throughput configurations, reserving only 5\% of HBM for runtime overhead and temporary allocations. Lower values (0.85--0.90) are appropriate for workloads requiring memory headroom for variable-length sequences or vision encoder activations.

\paragraph{Chunked Prefill Interaction.} Since vLLM V1 always enables chunked prefill, the \texttt{--max-num-\allowbreak{}batched-\allowbreak{}tokens} parameter also controls the prefill chunk size. Lower values (2,048) favor inter-token latency by processing smaller chunks and yielding back to decode more frequently. Higher values (32,768+) favor time-to-first-token (TTFT) and overall throughput by processing more prompt tokens per scheduler iteration, at the cost of slightly increased ITL for concurrent decode sequences.

\section{Experimental Methodology}
\label{sec:methodology}

This section describes the benchmark framework, test environment, and workload design used to evaluate large-scale LLM inference on AMD Instinct MI325X GPUs with vLLM.

\subsection{Benchmark Framework}
\label{subsec:benchmark-framework}

We developed a unified progressive benchmark that systematically evaluates inference serving performance across five phases of increasing intensity:

\begin{enumerate}
    \item \textbf{WARMUP} -- Initialize the model, warm GPU caches, and establish a single-request baseline.
    \item \textbf{BASELINE} -- Measure clean metrics at concurrency$=$1 to establish reference p99 latency.
    \item \textbf{SCALING} -- Sweep concurrency from 5 to 200 to characterize the throughput--latency trade-off.
    \item \textbf{STRESS} -- Test edge cases including long output generation (500+ tokens), long context prefill (4K--8K tokens), and multi-image vision workloads.
    \item \textbf{SATURATION} -- Find the breaking point by pushing concurrency from 150 to 1,000 concurrent requests.
\end{enumerate}

This progressive design ensures that each phase builds on prior results: the BASELINE p99 latency serves as the reference threshold for DEGRADED status detection (p99 $> 2\times$ baseline), and the SCALING phase throughput informs SATURATED status detection (throughput $< 1.05\times$ previous level).

\paragraph{Prompt Construction.}
Benchmarks use a fixed deterministic text prompt (a 4-sentence passage on machine learning concepts) repeated to reach the target input token count using a $\sim$4 characters-per-token heuristic. Prompts are not randomized, which may interact with vLLM's prefix caching optimizations. Vision benchmarks append image URLs to the same text prompt.

\paragraph{Warmup Protocol.}
Each benchmark run begins with a single warmup request (1 concurrent request, 100 input tokens, 50 output tokens) to trigger initial model compilation. No post-warmup cooldown is applied between the warmup and measurement phases. Warmup results are recorded but labeled separately as Phase~1 (WARMUP) data. Inter-phase cooldowns of 3--5 seconds are applied between subsequent measurement phases.

\paragraph{Concurrency Model.}
All concurrent requests are dispatched using Python's \texttt{Thread\-Pool\-Executor} with \texttt{max\_workers} set to the target concurrency level. Each worker issues a synchronous HTTP POST to the vLLM OpenAI-compatible \texttt{/v1/chat/\allowbreak{}completions} endpoint with a 300-second timeout. This approach accurately models real-world API gateway patterns where multiple client connections are served simultaneously.

\paragraph{Client-Side Measurement Validity.}
To rule out client-side bottlenecks, we benchmarked the client against a local echo server returning instant responses at concurrency levels from 100 to 2{,}000. The client sustained over 885 requests/second with 100\% success at all concurrency levels including 2{,}000 concurrent threads, with p99 latency under 25\,ms at concurrency levels ${\geq}$500 (matching the benchmark's saturation range). In contrast, vLLM benchmarks at saturation (${\sim}$500 concurrent) produce at most ${\sim}$19 requests/second for the fastest model (Qwen3-VL). The client's measured ceiling exceeds actual benchmark request rates by ${\sim}$50$\times$, confirming that the observed throughput saturation is server-side.

\paragraph{Metrics Collection.}
For each test, we collect the following metrics from successful responses:
\begin{itemize}
    \item \textbf{Latency percentiles}: p50, p95, and p99, computed from sorted response times.
    \item \textbf{Total throughput}: $(T_{\text{prompt}} + T_{\text{completion}}) / t_{\text{wall}}$ in tokens per second, where $T_{\text{prompt}}$ and $T_{\text{completion}}$ are cumulative token counts from the API usage response and $t_{\text{wall}}$ is the wall-clock duration.
    \item \textbf{Output throughput}: $T_{\text{completion}} / t_{\text{wall}}$ in tokens per second, reflecting pure generation speed.
    \item \textbf{Success rate}: fraction of requests returning HTTP~200 without client-side exceptions. This metric does not validate response content, JSON structure, or whether the model generated the requested number of output tokens. A response returning zero completion tokens with HTTP~200 would be counted as successful. Throughput metrics (tok/s) reflect actual tokens generated, providing an implicit output quality signal.
\end{itemize}

\paragraph{Status Classification.}
Each test result is classified into one of four status categories, summarized in Table~\ref{tab:status-indicators}.

\begin{table}[t]
\centering
\caption{Benchmark status indicators and their thresholds.}
\label{tab:status-indicators}
\begin{tabular}{@{}llp{5.5cm}@{}}
\toprule
\textbf{Status} & \textbf{Threshold} & \textbf{Interpretation} \\
\midrule
OK & Success $\geq 95\%$, latency normal & Normal operation \\
DEGRADED & p99 $> 2\times$ baseline p99 & Expected latency increase under concurrent load \\
SATURATED & Throughput $< 1.05\times$ previous & Throughput plateau reached \\
FAILING & Success $< 95\%$ & Requests are failing; reduce load \\
\bottomrule
\end{tabular}
\end{table}

Status labels use phase-relative criteria: the SCALING phase classifies requests as DEGRADED when p99 latency exceeds twice the single-request baseline, while the SATURATION phase uses throughput-relative classification only (SATURATED when throughput plateaus within 5\% of the previous level). The \texttt{baseline\_p99} threshold is only passed to the SCALING phase; consequently, DEGRADED status cannot trigger in the SATURATION or STRESS phases, and status labels are not directly comparable across phases.

\paragraph{Test Multipliers.}
The framework supports three intensity modes: \texttt{quick} ($0.5\times$ request count), \texttt{default} ($1\times$), and \texttt{thorough} ($3\times$). All stress test results reported in this paper use the \texttt{thorough} ($3\times$) multiplier to ensure statistical robustness, while validation results use the \texttt{quick} ($0.5\times$) multiplier for rapid correctness checks.

\paragraph{Multi-Run Reproducibility Protocol.}
To characterize measurement variance, we performed multiple independent benchmark runs per model using a standardized workload (100 requests, 2,048 input tokens, 512 output tokens, \texttt{quick} mode). Five runs were completed for all four models. Each run used a fresh vLLM server restart, including full model reloading and CUDA graph recapture, to ensure statistical independence between runs. Confidence intervals are computed using the $t$-distribution at the 95\% level with $\text{df} = n - 1$, appropriate for small sample sizes where the population variance is unknown.

\subsection{Test Environment}
\label{subsec:test-environment}

All experiments were conducted on a single server equipped with eight AMD Instinct MI325X GPUs. The complete hardware and software specifications are listed in Table~\ref{tab:test-environment}.

\begin{table}[t]
\centering
\caption{Test environment specifications.}
\label{tab:test-environment}
\begin{tabular}{@{}ll@{}}
\toprule
\textbf{Component} & \textbf{Specification} \\
\midrule
GPU & 8$\times$ AMD Instinct MI325X \\
Architecture & CDNA 3 (gfx942) \\
VRAM per GPU & 256\,GB HBM3e \\
Total VRAM & 2\,TB \\
Memory Bandwidth (per GPU) & 6.0\,TB/s \\
Aggregate Bandwidth & 48\,TB/s \\
FP16 Compute (per GPU) & 1,307\,TFLOPS \\
\midrule
ROCm & 6.4.2-120 \\
vLLM & 0.14.1 \\
Docker & 29.1.5 \\
RCCL & 2.26.6 \\
Python & 3.10+ (inside container) \\
Container (stable models) & \texttt{vllm/vllm-openai-rocm:latest}\textsuperscript{$\dagger$} \\
Container (Kimi-K2.5) & \texttt{rocm/vllm-dev:nightly}\textsuperscript{$\dagger$} \\
\midrule
System RAM & 256\,GB+ \\
CPU Cores & 64+ \\
NUMA Balancing & Disabled \\
\bottomrule
\end{tabular}
\end{table}

\textsuperscript{$\dagger$}Pinned digests: \texttt{vllm-openai-rocm} \texttt{sha256:2369\allowbreak{}00d5\allowbreak{}7300...}, \texttt{vllm-dev} \texttt{sha256:e8ce\allowbreak{}7f6d\allowbreak{}74a0...}. Full digests and model checkpoint revisions are listed in Table~\ref{tab:reproducibility-pins}.

\begin{table}[t]
\centering
\small
\caption{Pinned model checkpoint revisions and container image digests for reproducibility.}
\label{tab:reproducibility-pins}
\begin{tabular}{@{}l>{\raggedright\arraybackslash}p{5.8cm}@{}}
\toprule
\textbf{Artifact} & \textbf{Revision / Digest} \\
\midrule
DeepSeek-V3-0324 & \texttt{e9b33add7688} \\
Llama-3.1-405B-Instruct & \texttt{be673f326cab} \\
Qwen3-VL-235B-A22B & \texttt{710c13861be6} \\
Kimi-K2-Instruct & \texttt{1cbe779b5c9d} \\
\midrule
\texttt{vllm-openai-rocm} & \texttt{sha256:236900d57300\allowbreak{}1f7713e1526db1000dbaec6\allowbreak{}60d025645528ecc614d811\allowbreak{}d25cc5a} \\
\texttt{vllm-dev:nightly} & \texttt{sha256:e8ce7f6d74a0\allowbreak{}5dd11678920df707a5665d59\allowbreak{}98b438288da31406c9d655\allowbreak{}21a855} \\
\bottomrule
\end{tabular}
\end{table}

All models were served through Docker containers with ROCm device passthrough (\texttt{/dev/kfd}, \texttt{/dev/dri}), shared memory via \texttt{--ipc=host}, and model weights cached on the host filesystem. NUMA balancing was disabled (\texttt{kernel.numa\_\allowbreak{}balancing=0}) to prevent the OS from migrating memory pages across NUMA domains, which degrades multi-GPU communication latency.

\subsection{Workload Design}
\label{subsec:workload-design}

We evaluated four large language models spanning three distinct architectural families. Model specifications and architectural details are presented in Table~\ref{tab:model-specs} (Section~\ref{sec:architecture}).

\subsubsection{Validation Tests}
Validation tests use the \texttt{quick} ($0.5\times$) multiplier and serve two purposes: (1)~verifying functional correctness of the deployment (all API endpoints return valid responses) and (2)~establishing preliminary scaling curves for rapid comparison. Validation concurrency sweeps cover levels from 5 to 100, and saturation tests extend to 500 concurrent requests.

\subsubsection{Stress Tests}
Stress tests use the \texttt{thorough} ($3\times$) multiplier for comprehensive evaluation. The workload categories are:

\begin{itemize}
    \item \textbf{Concurrency scaling}: Sweep from 5 to 200 concurrent requests with 500-token input prompts and 200-token output generation, measuring throughput and latency at each level.
    \item \textbf{Long output generation}: 50 concurrent requests generating 500 output tokens each, measuring sustained generation throughput.
    \item \textbf{Long context prefill}: 25 concurrent requests with 4,000-token and 8,000-token input contexts, measuring prefill throughput.
    \item \textbf{Vision workloads} (Qwen3-VL, Kimi-K2.5 only): Multi-image requests with 3--5 images per request, high-concurrency vision at 100 concurrent, and sustained vision load at 50 concurrent over 450 requests.
    \item \textbf{Saturation}: Extreme concurrency at 150, 200, 300, 500, 750, and 1,000 concurrent requests to identify peak throughput and the saturation point.
\end{itemize}

\begin{table}[t]
\centering
\caption{Primary benchmark workloads by model. Vision-capable models were benchmarked with their native multimodal workload; text-only models used a text-only workload. Image tokens (processed by the vision encoder during prefill, not generated by the decoder) inflate total throughput for vision workloads, making total tok/s not directly comparable across workload types.}
\label{tab:primary-workloads}
\begin{tabular}{@{}llrrr@{}}
\toprule
\textbf{Model} & \textbf{Type} & \textbf{Input} & \textbf{Output} & \textbf{Images} \\
\midrule
Llama-3.1-405B & Text & 500 & 100 & 0 \\
DeepSeek V3.2 & Text & 500 & 100 & 0 \\
Qwen3-VL-235B & Vision & 100 & 200 & 1 \\
Kimi-K2.5 & Vision & 100 & 200 & 1 \\
\bottomrule
\end{tabular}
\end{table}

\paragraph{Workload comparability.} Each image adds approximately 1{,}000--1{,}400 tokens to the total token count (varying by model vision encoder), making \texttt{throughput\_total} not directly comparable across workload types. All cross-architecture comparisons in Section~\ref{sec:results} are therefore restricted to within-workload groups: text models (Llama-3.1-405B, DeepSeek~V3.2) and vision models (Qwen3-VL-235B, Kimi-K2.5).

\subsubsection{Architecture-Specific Configuration}
Each model required architecture-specific vLLM flags, summarized in Table~\ref{tab:model-flags}.

\begin{table}[t]
\centering
\caption{Architecture-specific vLLM configuration flags.}
\label{tab:model-flags}
\begin{tabular}{@{}l>{\raggedright\arraybackslash}p{7cm}@{}}
\toprule
\textbf{Model} & \textbf{Key Flags} \\
\midrule
DeepSeek V3.2 & \texttt{VLLM\_ROCM\_USE\_AITER=1}, \texttt{AITER\_ENABLE\_VSKIP=0}, \texttt{--block-size 1}, \texttt{--quantization fp8} \\
Llama-3.1-405B & \texttt{--quantization fp8}, \texttt{--max-model-len 32768} \\
Qwen3-VL-235B & \texttt{VLLM\_USE\_TRITON\_FLASH\_ATTN=0}, \texttt{--kv-offloading-backend native}, \texttt{--kv-offloading-size 64} \\
Kimi-K2.5 & \texttt{VLLM\_ROCM\_USE\_AITER=0}, \texttt{VLLM\_USE\_TRITON\_FLASH\_ATTN=0}, \texttt{--block-size 1}, \texttt{--tensor-parallel-size 4} \\
\bottomrule
\end{tabular}
\end{table}

Notable configuration constraints include: (1)~MLA-based models (DeepSeek~V3.2, Kimi-K2.5) require \texttt{--block-size~1} and do not support KV cache offloading or FP8 KV cache; (2)~Kimi-K2.5 is deployed with TP$=$4 (see Section~\ref{subsec:tensor-parallelism} for the rationale and limitations of this choice); and (3)~Qwen3-VL does not support FP8 quantization due to a model-architecture dimension constraint (intermediate MLP dimension of 4304 not divisible by 128, the FP8 block quantization block size).

\paragraph{Memory Measurement.}
All memory figures reported in this paper are \emph{per-GPU weight memory} as measured from vLLM's startup log line ``Model loading took X~GiB,'' which reports the weight memory allocated on a single GPU after model loading but before KV cache pre-allocation. These measurements were confirmed via dedicated memory verification benchmarks (Phase~3) that captured vLLM logs and cross-referenced with \texttt{rocm-smi} readings for all four models. Total GPU VRAM usage (visible via \texttt{rocm-smi}) is substantially higher because vLLM pre-allocates nearly all remaining VRAM for KV cache and CUDA graph capture buffers. For MoE models, reported parameter counts (e.g., 685B, 235B, 1T) represent architectural totals that sum all expert parameters independently. The actual stored parameters (and thus the GPU memory footprint) are smaller because MoE architectures share routing, embedding, and attention layers across experts.

\section{Results and Analysis}
\label{sec:results}

We present comprehensive benchmark results for four large language models deployed on 8$\times$ AMD Instinct MI325X GPUs using vLLM. All stress test results use the \texttt{thorough} ($3\times$) multiplier unless otherwise noted. Across all models, we processed over 18.9 million tokens across 17,406 requests with a 100\% HTTP-level success rate.

\subsection{Per-Model Performance}
\label{subsec:per-model}

Tables~\ref{tab:peak-performance-text} and~\ref{tab:peak-performance-vision} summarize peak performance metrics for each model, separated by workload type.

\begin{table}[t]
\centering
\caption{Peak performance summary: text-only workload (500-token input, 100-token output) on 8$\times$ MI325X. Peak throughput measured at the concurrency level that maximizes total tok/s during saturation testing. Saturation onset confirmed by fine-grained sweeps (500--1{,}000 in steps of 50, 3 runs per level).}
\label{tab:peak-performance-text}
\begin{tabular}{@{}lrrrrr@{}}
\toprule
\textbf{Model} & \textbf{Peak tok/s} & \textbf{Output tok/s} & \textbf{Peak Conc.} & \textbf{Sat.\ Point} & \textbf{Success} \\
\midrule
Llama-3.1-405B & 15,944 & 3,673 & 500  & 500 & 100\% \\
DeepSeek V3.2  & 15,343 & 1,239 & 500  & 500 & 100\% \\
\bottomrule
\end{tabular}
\end{table}

\begin{table}[t]
\centering
\caption{Peak performance summary: vision workload (100-token text + 1 image, 200-token output) on 8$\times$ MI325X. Total tok/s includes ${\sim}$1{,}000--1{,}400 image tokens per request and is not directly comparable to text-only results. Peak throughput measured at the concurrency level that maximizes total tok/s during saturation testing. Saturation onset confirmed by fine-grained sweeps (500--1{,}000 in steps of 50, 3 runs per level).}
\label{tab:peak-performance-vision}
\begin{tabular}{@{}lrrrrr@{}}
\toprule
\textbf{Model} & \textbf{Peak tok/s} & \textbf{Output tok/s} & \textbf{Peak Conc.} & \textbf{Sat.\ Point} & \textbf{Success} \\
\midrule
Qwen3-VL-235B  & 47,873 & 7,140 & 500  & 500 & 100\% \\
Kimi-K2.5      &  7,327 &   867 & 500  & 500 & 100\% \\
\bottomrule
\end{tabular}
\end{table}

\subsubsection{Qwen3-VL-235B-A22B (MoE + GQA)}

Qwen3-VL achieved the highest total throughput among the vision models tested, reaching 47,873 tok/s at 500 concurrent requests. Approximately 77\% of these tokens are image tokens processed by the vision encoder rather than output tokens generated by the LLM decoder (output throughput: 7,140 tok/s). This result is striking given that Qwen3-VL has the smallest total parameter count (235B) among the models tested. A key contributing factor is its MoE architecture with only 22B active parameters per token, combined with Grouped-Query Attention (GQA), which enables exceptionally efficient batching.

Table~\ref{tab:qwen3-scaling} shows the concurrency scaling profile.

\begin{table}[t]
\centering
\caption{Qwen3-VL-235B concurrency scaling results (stress test, $3\times$ multiplier).}
\label{tab:qwen3-scaling}
\begin{tabular}{@{}rrrrl@{}}
\toprule
\textbf{Conc.} & \textbf{Total tok/s} & \textbf{Output tok/s} & \textbf{p99 Latency} & \textbf{Status} \\
\midrule
10  & 3,586  & 535   & 3.78s  & DEGRADED \\
25  & 7,728  & 1,153 & 4.36s  & DEGRADED \\
50  & 13,489 & 2,012 & 5.09s  & DEGRADED \\
100 & 21,567 & 3,217 & 6.33s  & DEGRADED \\
200 & 26,674 & 3,978 & 8.92s  & DEGRADED \\
\bottomrule
\end{tabular}
\end{table}

The model exhibited strong sublinear scaling up to 200 concurrent requests, with throughput increasing 13.6$\times$ from 5 to 200 concurrent (a 40$\times$ concurrency increase). Stress testing revealed strong prefill performance: 14,193 tok/s for 4K-token contexts and efficient multi-image processing (3--5 images per request handled without failures). Long output generation reached 1,959 output tok/s.

\subsubsection{Llama-3.1-405B-Instruct (Dense + GQA)}

As the only dense model in our evaluation, Llama-3.1-405B provides a reference point for understanding MoE benefits. It achieved a peak throughput of 15,944 tok/s at 500 concurrent requests, comparable to DeepSeek~V3.2 (15,343 tok/s) despite having an order of magnitude more active parameters (405B vs.\ 37B), consistent with MoE sparsity's potential to match dense model throughput with far fewer active parameters.

Table~\ref{tab:llama-scaling} shows the scaling profile.

\begin{table}[t]
\centering
\caption{Llama-3.1-405B concurrency scaling results (stress test, $3\times$ multiplier).}
\label{tab:llama-scaling}
\begin{tabular}{@{}rrrrl@{}}
\toprule
\textbf{Conc.} & \textbf{Total tok/s} & \textbf{Output tok/s} & \textbf{p99 Latency} & \textbf{Status} \\
\midrule
5   & 423    & 153   & 6.15s  & DEGRADED \\
10  & 851    & 322   & 6.23s  & DEGRADED \\
25  & 1,953  & 738   & 6.80s  & DEGRADED \\
50  & 3,428  & 1,296 & 7.75s  & DEGRADED \\
100 & 5,254  & 1,986 & 10.08s & DEGRADED \\
150 & 6,927  & 2,619 & 11.45s & DEGRADED \\
\bottomrule
\end{tabular}
\end{table}

Llama-3.1-405B demonstrated the most predictable scaling behavior, with p99 latency increasing only 2.4$\times$ (from 6.15s to 14.49s under the scaling workload of 500-token input, 200-token output) across the full concurrency range. Stress tests showed efficient long-context prefill at 8,240 tok/s for 4K contexts and 6,794 tok/s for 8K contexts. Long output generation achieved 1,224 output tok/s.

\subsubsection{DeepSeek~V3.2 (MoE + MLA)}

DeepSeek~V3.2 reached a peak throughput of 15,343 tok/s at 500 concurrent requests, comparable to Llama-3.1-405B despite having 685B total parameters (37B active). The MLA attention mechanism enables competitive throughput at scale but exhibits earlier throughput saturation during the scaling phase relative to GQA-based models.

Table~\ref{tab:deepseek-scaling} presents the scaling results.

\begin{table}[t]
\centering
\caption{DeepSeek~V3.2 concurrency scaling results (stress test, $3\times$ multiplier).}
\label{tab:deepseek-scaling}
\begin{tabular}{@{}rrrrl@{}}
\toprule
\textbf{Conc.} & \textbf{Total tok/s} & \textbf{Output tok/s} & \textbf{p99 Latency} & \textbf{Status} \\
\midrule
5   & 461    & 106   & 8.19s  & OK \\
10  & 1,235  & 162   & 8.12s  & OK \\
25  & 3,013  & 477   & 8.72s  & DEGRADED \\
50  & 4,607  & 802   & 11.34s & DEGRADED \\
100 & 7,160  & 981   & 13.75s & DEGRADED \\
200 & 7,266  & 1,045 & 13.95s & DEGRADED \\
\bottomrule
\end{tabular}
\end{table}

Scaling throughput peaked at 7,266 tok/s at 200 concurrent during the scaling phase, with the saturation-phase tests pushing this to 15,343 tok/s at 500 concurrent. Stress tests showed 4,274 tok/s for 4K-context workloads and 3,372 tok/s for 8K-context workloads, with long output generation at 230 output tok/s. The p99 latency increased moderately from 8s to 14s across the scaling range.

\subsubsection{Kimi-K2.5 (MoE + MLA, 1T Parameters)}

Kimi-K2.5 is the largest model in our evaluation at 1 trillion total parameters with 32B active per token. It achieved a peak throughput of 7,327 tok/s at 500 concurrent requests. As the only model requiring TP$=$4 (due to MLA attention head constraints) and running without AITER acceleration, its throughput is lower than models with full kernel optimization support.

Table~\ref{tab:kimi-scaling} presents the concurrency scaling profile.

\begin{table}[t]
\centering
\caption{Kimi-K2.5 concurrency scaling results (stress test, $3\times$ multiplier).\protect\footnote{The concurrency~1 entry (587~tok/s) uses the vision-enabled baseline workload. The text-only baseline throughput is 151~tok/s. The status reversal from DEGRADED (concurrency 10--200) to OK (concurrency 500) reflects different classification criteria: the scaling phase labels requests DEGRADED when p99 latency exceeds twice the single-request baseline, while the saturation phase uses throughput-relative classification only (see Tables~\ref{tab:saturation-detail-text} and~\ref{tab:saturation-detail-vision}).}}
\label{tab:kimi-scaling}
\begin{tabular}{@{}rrrrl@{}}
\toprule
\textbf{Conc.} & \textbf{Total tok/s} & \textbf{Output tok/s} & \textbf{p99 Latency} & \textbf{Status} \\
\midrule
1   & 587    & 37    & 2.73s   & OK \\
10  & 670    & 79    & 25.40s  & DEGRADED \\
25  & 896    & 106   & 47.56s  & DEGRADED \\
50  & 1,632  & 193   & 52.02s  & DEGRADED \\
100 & 2,656  & 314   & 63.86s  & DEGRADED \\
200 & 3,754  & 444   & 74.89s  & DEGRADED \\
500 & 7,327  & 867   & 103.34s & OK \\
\bottomrule
\end{tabular}
\end{table}

The model exhibited excellent linear scaling from 5 to 200 concurrent requests, with consistent throughput from 500 to 1,000 concurrent ($\sim$7,300 tok/s). Vision workloads performed well: multi-image processing (3 images) reached 1,475 tok/s with 77.2s p99 latency, while sustained vision load (50 concurrent, 450 requests) maintained 836 tok/s over 17 minutes. Critically, the system maintained a 100\% success rate at all concurrency levels including 1,000 concurrent requests.

\subsection{Architecture Comparison}
\label{subsec:arch-comparison}

Our evaluation spans three architectural families, enabling direct comparison of their inference characteristics. Tables~\ref{tab:arch-comparison-text} and~\ref{tab:arch-comparison-vision} summarize the key differences, separated by workload type.

\begin{table}[t]
\centering
\caption{Architecture comparison: text-only workload. Throughput ratios are relative to Llama-3.1-405B (Dense+GQA) as the baseline.}
\label{tab:arch-comparison-text}
\begin{tabular}{@{}lcrrr@{}}
\toprule
\textbf{Architecture} & \textbf{Model} & \textbf{Active Params} & \textbf{Peak tok/s} & \textbf{Ratio} \\
\midrule
Dense + GQA & Llama-3.1-405B & 405B & 15,944 & 1.00$\times$ \\
MoE + MLA   & DeepSeek V3.2  & 37B  & 15,343 & 0.96$\times$ \\
\bottomrule
\end{tabular}
\end{table}

\begin{table}[t]
\centering
\caption{Architecture comparison: vision workload. Throughput ratios are relative to Kimi-K2.5 (MoE+MLA) as the baseline.}
\label{tab:arch-comparison-vision}
\begin{tabular}{@{}lcrrr@{}}
\toprule
\textbf{Architecture} & \textbf{Model} & \textbf{Active Params} & \textbf{Peak tok/s} & \textbf{Ratio} \\
\midrule
MoE + GQA   & Qwen3-VL-235B  & 22B  & 47,873 & 6.53$\times$ \\
MoE + MLA   & Kimi-K2.5      & 32B  & 7,327  & 1.00$\times$ \\
\bottomrule
\end{tabular}
\end{table}

\paragraph{Text workload: Dense+GQA vs.\ MoE+MLA.} Llama-3.1-405B (Dense+GQA, 405B active) and DeepSeek~V3.2 (MoE+MLA, 37B active) achieve nearly identical total throughput on the text workload (15,944 vs.\ 15,343 tok/s), despite DeepSeek having only 9\% of Llama's active parameters. This parity suggests that DeepSeek's MoE sparsity advantage is offset by MLA's constraints on the current ROCm stack, including the block-size-1 requirement and inability to use KV cache offloading. However, Llama achieves substantially higher output throughput (3,673 vs.\ 1,239 tok/s), indicating differences in decoder generation speed.

\paragraph{Vision workload: MoE+GQA vs.\ MoE+MLA.} Among the vision models, Qwen3-VL-235B (MoE+GQA, 22B active) achieves 6.5$\times$ the total throughput of Kimi-K2.5 (MoE+MLA, 32B active). Even in output throughput (7,140 vs.\ 867 tok/s, an 8.2$\times$ ratio), Qwen3-VL substantially outperforms Kimi-K2.5. This gap reflects several compounding factors: Qwen3-VL's fewer active parameters (22B vs.\ 32B), its GQA attention enabling KV cache offloading and standard block sizes, AITER-enabled acceleration (vs.\ disabled for Kimi), and higher tensor parallelism degree (TP$=$8 vs.\ TP$=$4).

\paragraph{Active Parameters and Throughput.}
A key finding is that active parameters per token are more consistently associated with throughput than total parameters across the models tested. Tables~\ref{tab:throughput-per-param-text} and~\ref{tab:throughput-per-param-vision} quantify this relationship.

\begin{table}[t]
\centering
\caption{Throughput normalized by active parameter count: text-only workload. Higher values indicate more efficient utilization of active compute.}
\label{tab:throughput-per-param-text}
\begin{tabular}{@{}lrrr@{}}
\toprule
\textbf{Model} & \textbf{Active (B)} & \textbf{Peak tok/s} & \textbf{tok/s per B} \\
\midrule
Llama-3.1-405B & 405 & 15,944 & 39 \\
DeepSeek V3.2  & 37  & 15,343 & 415 \\
\bottomrule
\end{tabular}
\end{table}

\begin{table}[t]
\centering
\caption{Throughput normalized by active parameter count: vision workload. Higher values indicate more efficient utilization of active compute. Vision values include image tokens in the numerator and are not directly comparable to text-workload values.}
\label{tab:throughput-per-param-vision}
\begin{tabular}{@{}lrrr@{}}
\toprule
\textbf{Model} & \textbf{Active (B)} & \textbf{Peak tok/s} & \textbf{tok/s per B} \\
\midrule
Qwen3-VL-235B  & 22  & 47,873 & 2,176 \\
Kimi-K2.5      & 32  & 7,327  & 229 \\
\bottomrule
\end{tabular}
\end{table}

Within the text workload, DeepSeek~V3.2 achieves 415 tok/s per billion active parameters, 10.6$\times$ higher than dense Llama-3.1-405B (39 tok/s/B), quantifying the throughput efficiency of MoE sparsity per active parameter. Within the vision workload, Qwen3-VL achieves 2,176 tok/s/B, 9.5$\times$ higher than Kimi-K2.5 (229 tok/s/B); the Kimi-K2.5 gap is partially attributable to its disabled AITER acceleration and TP$=$4 constraint. The substantially higher absolute tok/s/B values for vision models reflect the inclusion of image tokens in the numerator and are not directly comparable to the text-workload values.

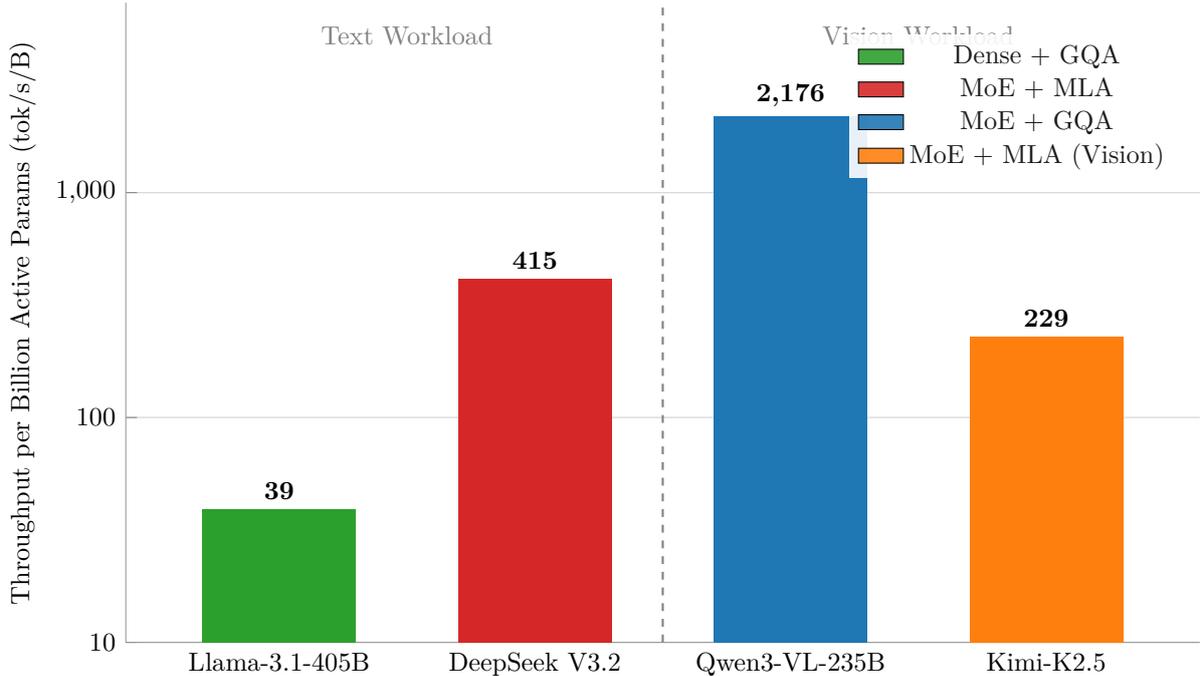
\begin{figure}[t]
\centering
\resizebox{\columnwidth}{!}{
\begin{tikzpicture}
\begin{axis}[
  width=\columnwidth,
  height=0.64\columnwidth,
  ylabel={Throughput per Billion Active Params (tok/s/B)},
  ymin=10, ymax=7000,
  ymode=log,
  log basis y=10,
  xmin=-0.6, xmax=3.6,
  xtick={0,1,2,3},
  xticklabels={Llama-3.1-405B, DeepSeek V3.2, Qwen3-VL-235B, Kimi-K2.5},
  xticklabel style={font=\small},
  grid=major,
  grid style={gray!30, line width=0.5pt},
  ymajorgrids=true,
  xmajorgrids=false,
  ytick={10,100,1000},
  yticklabels={10, 100, {1,000}},
  tick label style={font=\small},
  label style={font=\small},
  axis line style={gray!80},
  axis x line*=bottom,
  axis y line*=left,
  legend style={font=\small, at={(0.98,0.95)}, anchor=north east, draw=none, fill opacity=0.9},
  clip=false,
]

\draw[gray, dashed, line width=0.8pt] (axis cs:1.5,10) -- (axis cs:1.5,7000);

\node[font=\small, text=gray] at (axis cs:0.5,5000) {Text Workload};
\node[font=\small, text=gray] at (axis cs:2.5,5000) {Vision Workload};


\fill[fill={rgb,255:red,44;green,160;blue,44}]
  (axis cs:-0.3,10) rectangle (axis cs:0.3,39);
\node[above, font=\small\bfseries] at (axis cs:0,39) {39};

\fill[fill={rgb,255:red,214;green,39;blue,40}]
  (axis cs:0.7,10) rectangle (axis cs:1.3,415);
\node[above, font=\small\bfseries] at (axis cs:1,415) {415};

\fill[fill={rgb,255:red,31;green,119;blue,180}]
  (axis cs:1.7,10) rectangle (axis cs:2.3,2176);
\node[above, font=\small\bfseries] at (axis cs:2,2176) {2,176};

\fill[fill={rgb,255:red,255;green,127;blue,14}]
  (axis cs:2.7,10) rectangle (axis cs:3.3,229);
\node[above, font=\small\bfseries] at (axis cs:3,229) {229};

\addlegendimage{area legend, fill={rgb,255:red,44;green,160;blue,44}}
\addlegendentry{Dense + GQA}
\addlegendimage{area legend, fill={rgb,255:red,214;green,39;blue,40}}
\addlegendentry{MoE + MLA}
\addlegendimage{area legend, fill={rgb,255:red,31;green,119;blue,180}}
\addlegendentry{MoE + GQA}
\addlegendimage{area legend, fill={rgb,255:red,255;green,127;blue,14}}
\addlegendentry{MoE + MLA (Vision)}

\end{axis}
\end{tikzpicture}}
\caption{Throughput normalized by active parameter count (tok/s per billion active parameters), grouped by workload type. Left pair: text-only workload; right pair: vision workload. Values are not directly comparable across workload types because vision total tok/s includes image tokens. Y-axis uses logarithmic scale. Data from primary stress-test benchmark ($3\times$ multiplier).}
\label{fig:throughput-per-param}
\end{figure}

\subsection{Scaling and Saturation Analysis}
\label{subsec:scaling-saturation}

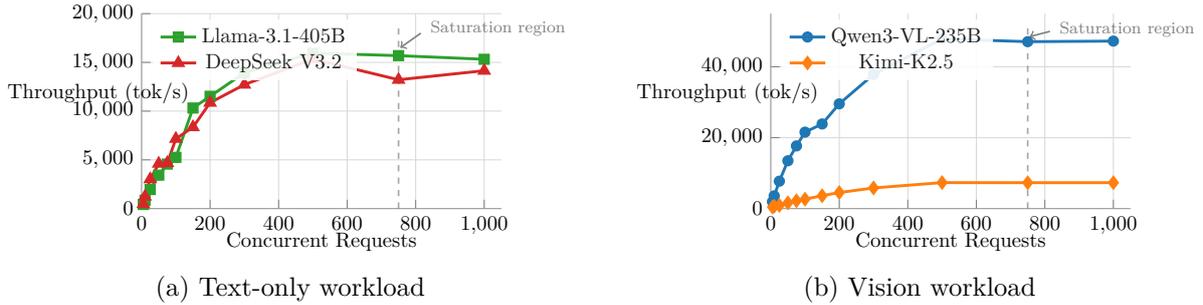
\begin{figure}[t]
\centering
\begin{minipage}[t]{0.48\columnwidth}
\centering
\resizebox{\textwidth}{!}{
\begin{tikzpicture}
\begin{axis}[
  width=\columnwidth,
  height=0.64\columnwidth,
  xlabel={Concurrent Requests},
  ylabel={Throughput (tok/s)},
  xmin=0, xmax=1050,
  ymin=0, ymax=20000,
  grid=major,
  grid style={gray!30, line width=0.5pt},
  legend pos=north west,
  legend style={font=\small, fill opacity=0.9, draw=none},
  every axis x label/.style={at={(0.5,-0.08)}, anchor=north},
  every axis y label/.style={at={(-0.12,0.5)}, anchor=south},
  scaled y ticks=false,
  yticklabel={\pgfmathprintnumber[fixed,1000 sep={,}]{\tick}},
  tick label style={font=\small},
  label style={font=\small},
  axis line style={gray!80},
  every outer x axis line/.append style={gray!80},
  every outer y axis line/.append style={gray!80},
  axis x line*=bottom,
  axis y line*=left,
  clip=false,
]


\addplot[color={rgb,255:red,44;green,160;blue,44}, mark=square*, mark size=2pt, line width=1.4pt]
  coordinates {
    (5,422.8) (10,851.2) (25,1952.6) (50,3427.8) (75,4564.7)
    (100,5254.0) (150,10319.7) (200,11519.1) (300,13937.1)
    (500,15943.6) (750,15692.8) (1000,15319.1)
  };
\addlegendentry{Llama-3.1-405B}

\addplot[color={rgb,255:red,214;green,39;blue,40}, mark=triangle*, mark size=2.5pt, line width=1.4pt]
  coordinates {
    (5,461.3) (10,1234.5) (25,3013.0) (50,4606.7) (75,4676.6)
    (100,7159.7) (150,8355.2) (200,10864.0) (300,12719.5)
    (500,15342.9) (750,13218.1) (1000,14147.8)
  };
\addlegendentry{DeepSeek V3.2}

\draw[gray, dashed, line width=0.8pt, opacity=0.7] (axis cs:750,0) -- (axis cs:750,19000);
\node[anchor=west, font=\scriptsize, text=gray] at (axis cs:820, 18500) {Saturation region};
\draw[->, gray, line width=0.8pt] (axis cs:815, 18000) -- (axis cs:755, 16500);

\end{axis}
\end{tikzpicture}}
\subcaption{Text-only workload}
\end{minipage}
\hfill
\begin{minipage}[t]{0.48\columnwidth}
\centering
\resizebox{\textwidth}{!}{
\begin{tikzpicture}
\begin{axis}[
  width=\columnwidth,
  height=0.64\columnwidth,
  xlabel={Concurrent Requests},
  ylabel={Throughput (tok/s)},
  xmin=0, xmax=1050,
  ymin=0, ymax=55000,
  grid=major,
  grid style={gray!30, line width=0.5pt},
  legend pos=north west,
  legend style={font=\small, fill opacity=0.9, draw=none},
  every axis x label/.style={at={(0.5,-0.08)}, anchor=north},
  every axis y label/.style={at={(-0.12,0.5)}, anchor=south},
  scaled y ticks=false,
  yticklabel={\pgfmathprintnumber[fixed,1000 sep={,}]{\tick}},
  tick label style={font=\small},
  label style={font=\small},
  axis line style={gray!80},
  every outer x axis line/.append style={gray!80},
  every outer y axis line/.append style={gray!80},
  axis x line*=bottom,
  axis y line*=left,
  clip=false,
]


\addplot[color={rgb,255:red,31;green,119;blue,180}, mark=*, mark size=2pt, line width=1.4pt]
  coordinates {
    (5,1955.6) (10,3586.4) (25,7728.1) (50,13489.1) (75,17685.0)
    (100,21566.9) (150,23882.1) (200,29557.2) (300,37849.5)
    (500,47872.8) (750,47091.8) (1000,47252.5)
  };
\addlegendentry{Qwen3-VL-235B}

\addplot[color={rgb,255:red,255;green,127;blue,14}, mark=diamond*, mark size=2.5pt, line width=1.4pt]
  coordinates {
    (5,429.6) (10,669.9) (25,895.9) (50,1632.0) (75,2213.1)
    (100,2656.2) (150,3627.7) (200,4527.8) (300,5819.8)
    (500,7327.2) (750,7304.0) (1000,7309.4)
  };
\addlegendentry{Kimi-K2.5}

\draw[gray, dashed, line width=0.8pt, opacity=0.7] (axis cs:750,0) -- (axis cs:750,52000);
\node[anchor=west, font=\scriptsize, text=gray] at (axis cs:820, 50600) {Saturation region};
\draw[->, gray, line width=0.8pt] (axis cs:815, 50000) -- (axis cs:755, 48500);

\end{axis}
\end{tikzpicture}}
\subcaption{Vision workload}
\end{minipage}
\caption{Throughput scaling as a function of concurrent requests, separated by workload type. Both panels show saturation at approximately 500 concurrent requests and flat throughput through 1{,}000. All models maintain 100\% success rates. Data from primary stress-test benchmark ($3\times$ multiplier).}
\label{fig:scaling-curves}
\end{figure}

\paragraph{Linear Scaling Region.}
All four models exhibit sublinear throughput scaling up to 200 concurrent requests within their respective workloads. Among text models, Llama-3.1-405B scales to 6,927 tok/s at 150 concurrent before slight degradation at 200, while DeepSeek~V3.2 shows strong scaling from 461 tok/s at 5 concurrent to 7,266 tok/s at 200 concurrent. Among vision models, Qwen3-VL scales most efficiently, achieving a 13.6$\times$ throughput increase from 5 to 200 concurrent, while Kimi-K2.5 achieves a 6.4$\times$ increase from baseline to 200 concurrent (587 to 3,754 tok/s).

\paragraph{Saturation Behavior.}
A notable finding is that within each workload type, all models reach peak throughput at approximately 500 concurrent requests and plateau through 1{,}000. A fine-grained concurrency sweep from 500 to 1{,}000 in steps of 50 (three independent runs per level, 200 requests each) confirms flat throughput across this entire range: Qwen3-VL varies by 0.65\%, Kimi-K2.5 by 1.2\%, DeepSeek~V3.2 by 1.9\%, and Llama-3.1-405B by 2.0\% (relative range $(\max{-}\min)/\text{mean}$ across per-level mean throughputs, three runs per concurrency level), with no performance cliff or non-monotonic behavior. Tables~\ref{tab:saturation-detail-text} and~\ref{tab:saturation-detail-vision} detail the saturation-phase behavior.

\begin{table}[t]
\centering
\caption{Saturation-phase throughput (tok/s) and p99 latency at extreme concurrency levels: text-only workload.\protect\footnote{Status labels use phase-relative criteria: the SCALING phase classifies requests as DEGRADED when p99 latency exceeds twice the single-request baseline, while the SATURATION phase uses throughput-relative classification only (SATURATED when throughput plateaus within 5\% of the previous level). Consequently, status labels are not directly comparable across phases. Note that DeepSeek~V3.2 at concurrency 1000 shows OK status despite lower throughput than at 500; the non-monotonic recovery (13,218 $\rightarrow$ 14,148 tok/s) exceeds the 5\% growth threshold, preventing SATURATED classification.}}
\label{tab:saturation-detail-text}
\begin{tabular}{@{}l rr rr@{}}
\toprule
\textbf{Conc.} & \multicolumn{2}{c}{\textbf{Llama-405B}} & \multicolumn{2}{c}{\textbf{DeepSeek}} \\
\cmidrule(lr){2-3} \cmidrule(lr){4-5}
 & tok/s & p99 & tok/s & p99 \\
\midrule
150  & 10,320 & 6.15s  & 8,355  & 5.77s  \\
200  & 11,519 & 7.35s  & 10,864 & 5.73s  \\
300  & 13,937 & 9.09s  & 12,719 & 7.35s  \\
500  & 15,944 & 11.78s & 15,343 & 9.08s  \\
750  & 15,693 & 12.01s & 13,218 & 10.84s \\
1000 & 15,319 & 12.28s & 14,148 & 9.99s  \\
\bottomrule
\end{tabular}
\end{table}

\begin{table}[t]
\centering
\caption{Saturation-phase throughput (tok/s) and p99 latency at extreme concurrency levels: vision workload. Total tok/s includes image tokens processed by the vision encoder.}
\label{tab:saturation-detail-vision}
\begin{tabular}{@{}l rr rr@{}}
\toprule
\textbf{Conc.} & \multicolumn{2}{c}{\textbf{Qwen3-VL}} & \multicolumn{2}{c}{\textbf{Kimi-K2.5}} \\
\cmidrule(lr){2-3} \cmidrule(lr){4-5}
 & tok/s & p99 & tok/s & p99 \\
\midrule
150  & 23,557 & 8.50s  & 3,628  & 69.74s  \\
200  & 29,557 & 8.96s  & 4,528  & 74.44s  \\
300  & 37,850 & 10.50s & 5,820  & 86.81s  \\
500  & 47,873 & 12.37s & 7,327  & 103.34s \\
750  & 47,092 & 12.53s & 7,304  & 103.66s \\
1000 & 47,252 & 12.50s & 7,309  & 103.64s \\
\bottomrule
\end{tabular}
\end{table}

\paragraph{Reliability Under Extreme Load.}
The most significant operational finding is that all four models maintained a \textbf{100\% HTTP-level success rate} (every request returned HTTP~200 with a valid response structure; this metric does not validate output quality or token count correctness) across all concurrency levels, including 1,000 simultaneous requests. No request failures were observed even at saturation, indicating that vLLM's continuous batching and scheduling mechanisms gracefully handle overload by queuing excess requests rather than rejecting them.

\paragraph{Latency Trade-offs.}
The latency cost of high-concurrency throughput is manageable. Llama-3.1-405B showed the most predictable latency growth: p99 increased from 6.15s at 5~concurrent to 14.49s at 200~concurrent (2.4$\times$) under the scaling workload (500-token input, 200-token output; note that the saturation-phase workload in Table~\ref{tab:saturation-detail-text} uses a shorter 100-token output). Qwen3-VL p99 latency increased only 2.6$\times$ from 5 to 200~concurrent despite a 13.6$\times$ throughput gain, representing the best throughput-per-latency-cost ratio among all models. DeepSeek~V3.2 p99 latency increased moderately from 8s to 14s across the scaling range.

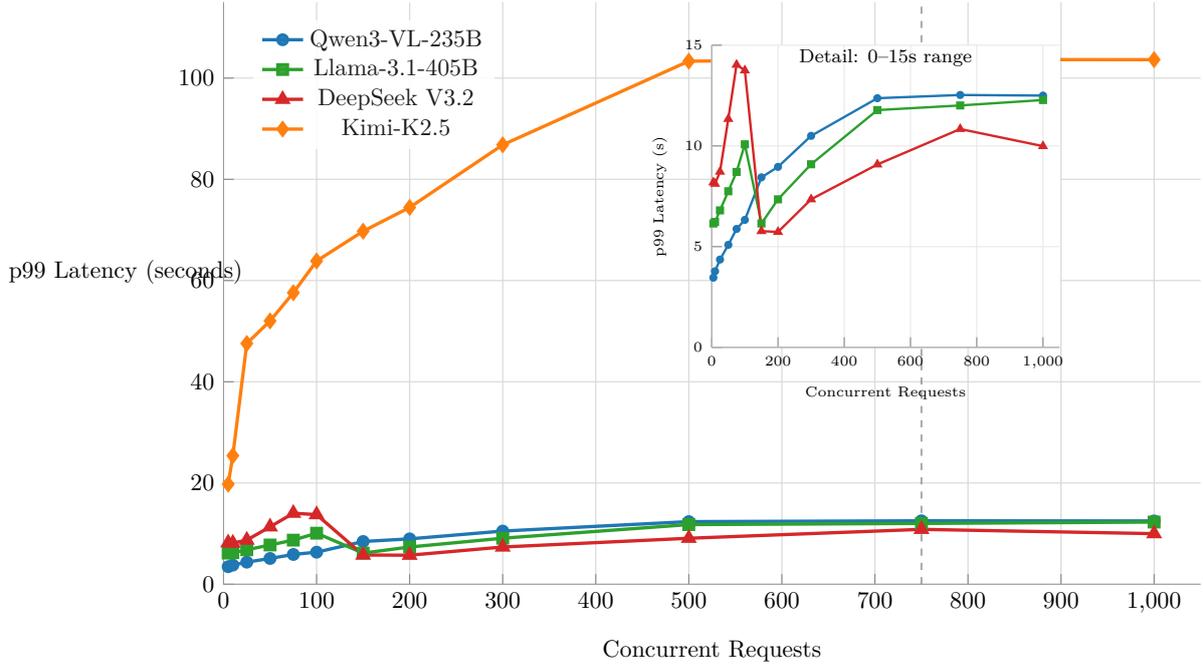
\begin{figure}[t]
\centering
\resizebox{\columnwidth}{!}{
\begin{tikzpicture}
\begin{axis}[
  width=\columnwidth,
  height=0.64\columnwidth,
  xlabel={Concurrent Requests},
  ylabel={p99 Latency (seconds)},
  xmin=0, xmax=1050,
  ymin=0, ymax=115,
  grid=major,
  grid style={gray!30, line width=0.5pt},
  legend pos=north west,
  legend style={font=\small, fill opacity=0.9, draw=none},
  every axis x label/.style={at={(0.5,-0.08)}, anchor=north},
  every axis y label/.style={at={(-0.1,0.5)}, anchor=south},
  tick label style={font=\small},
  label style={font=\small},
  axis line style={gray!80},
  axis x line*=bottom,
  axis y line*=left,
]

\addplot[color={rgb,255:red,31;green,119;blue,180}, mark=*, mark size=2pt, line width=1.4pt]
  coordinates {
    (5,3.46) (10,3.78) (25,4.36) (50,5.09) (75,5.88)
    (100,6.33) (150,8.44) (200,8.96) (300,10.50)
    (500,12.37) (750,12.53) (1000,12.50)
  };
\addlegendentry{Qwen3-VL-235B}

\addplot[color={rgb,255:red,44;green,160;blue,44}, mark=square*, mark size=2pt, line width=1.4pt]
  coordinates {
    (5,6.15) (10,6.23) (25,6.80) (50,7.75) (75,8.71)
    (100,10.08) (150,6.15) (200,7.35) (300,9.09)
    (500,11.78) (750,12.01) (1000,12.28)
  };
\addlegendentry{Llama-3.1-405B}

\addplot[color={rgb,255:red,214;green,39;blue,40}, mark=triangle*, mark size=2.5pt, line width=1.4pt]
  coordinates {
    (5,8.19) (10,8.12) (25,8.72) (50,11.34) (75,14.03)
    (100,13.75) (150,5.77) (200,5.73) (300,7.35)
    (500,9.08) (750,10.84) (1000,9.99)
  };
\addlegendentry{DeepSeek V3.2}

\addplot[color={rgb,255:red,255;green,127;blue,14}, mark=diamond*, mark size=2.5pt, line width=1.4pt]
  coordinates {
    (5,19.77) (10,25.40) (25,47.56) (50,52.02) (75,57.59)
    (100,63.86) (150,69.74) (200,74.44) (300,86.81)
    (500,103.34) (750,103.66) (1000,103.64)
  };
\addlegendentry{Kimi-K2.5}

\draw[gray, dashed, line width=0.8pt, opacity=0.7] (axis cs:750,0) -- (axis cs:750,115);

\end{axis}

\begin{axis}[
  at={(0.45\columnwidth, 0.22\columnwidth)},
  width=0.42\columnwidth,
  height=0.38\columnwidth,
  xlabel={\tiny Concurrent Requests},
  ylabel={\tiny p99 Latency (s)},
  title={\scriptsize Detail: 0--15s range},
  title style={at={(0.5,0.97)}, anchor=north, font=\scriptsize},
  xmin=0, xmax=1050,
  ymin=0, ymax=15,
  grid=major,
  grid style={gray!20, line width=0.3pt},
  tick label style={font=\tiny},
  label style={font=\tiny},
  axis line style={gray!60},
  axis x line*=bottom,
  axis y line*=left,
  axis background/.style={fill=white},
  every outer x axis line/.append style={line width=0.6pt},
  every outer y axis line/.append style={line width=0.6pt},
]

\addplot[color={rgb,255:red,31;green,119;blue,180}, mark=*, mark size=1.2pt, line width=1.0pt]
  coordinates {
    (5,3.46) (10,3.78) (25,4.36) (50,5.09) (75,5.88)
    (100,6.33) (150,8.44) (200,8.96) (300,10.50)
    (500,12.37) (750,12.53) (1000,12.50)
  };

\addplot[color={rgb,255:red,44;green,160;blue,44}, mark=square*, mark size=1.2pt, line width=1.0pt]
  coordinates {
    (5,6.15) (10,6.23) (25,6.80) (50,7.75) (75,8.71)
    (100,10.08) (150,6.15) (200,7.35) (300,9.09)
    (500,11.78) (750,12.01) (1000,12.28)
  };

\addplot[color={rgb,255:red,214;green,39;blue,40}, mark=triangle*, mark size=1.5pt, line width=1.0pt]
  coordinates {
    (5,8.19) (10,8.12) (25,8.72) (50,11.34) (75,14.03)
    (100,13.75) (150,5.77) (200,5.73) (300,7.35)
    (500,9.08) (750,10.84) (1000,9.99)
  };

\end{axis}
\end{tikzpicture}}
\caption{p99 latency as a function of concurrent requests. Text models (Llama-3.1-405B, DeepSeek~V3.2) use a 500-token input / 100-token output workload; vision models (Qwen3-VL, Kimi-K2.5) use a 100-token input + 1 image / 200-token output workload. Latency values are not directly comparable across workload types. All models show sublinear latency growth: throughput increases faster than latency, yielding positive scaling efficiency at all tested concurrency levels. Inset shows the 0--15s range for Qwen3-VL, Llama-3.1-405B, and DeepSeek~V3.2 (Kimi-K2.5 latencies of 25--103s compress these curves in the main plot). Data from primary stress-test benchmark ($3\times$ multiplier).}
\label{fig:latency-curves}
\end{figure}

\subsection{Measurement Reproducibility}
\label{subsec:reproducibility}

To characterize measurement variance, we conducted multiple independent benchmark runs per model using a standardized workload (100 requests, 2,048 input tokens, 512 output tokens). Each run used a fresh vLLM server restart to ensure independence. Five runs were completed for all four models. Confidence intervals use the $t$-distribution at the 95\% level.

\begin{table}[t]
\centering
\caption{Multi-run reproducibility statistics (peak throughput at the concurrency level maximizing mean tok/s). CoV is the coefficient of variation (stdev/mean). This workload differs from the primary benchmarks (Tables~\ref{tab:peak-performance-text} and~\ref{tab:peak-performance-vision}), which use 17,406 requests with a $3\times$ multiplier; values are not directly comparable. Confidence intervals use the $t$-distribution at 95\%.}
\label{tab:reproducibility}
\begin{tabular}{@{}lrrrrr@{}}
\toprule
\textbf{Model} & \textbf{$n$} & \textbf{Peak Conc.} & \textbf{Mean tok/s} & \textbf{CI\textsubscript{95}} & \textbf{CoV} \\
\midrule
Qwen3-VL-235B  & 5 & 1000 & 11,218 & $\pm$32   & 0.2\% \\
Llama-3.1-405B & 5 & 750  & 6,808  & $\pm$336  & 4.0\% \\
DeepSeek V3.2  & 5 & 1000 & 5,786  & $\pm$842  & 11.7\% \\
Kimi-K2.5      & 5 & 1000 & 952    & $\pm$4    & 0.4\% \\
\bottomrule
\end{tabular}
\end{table}

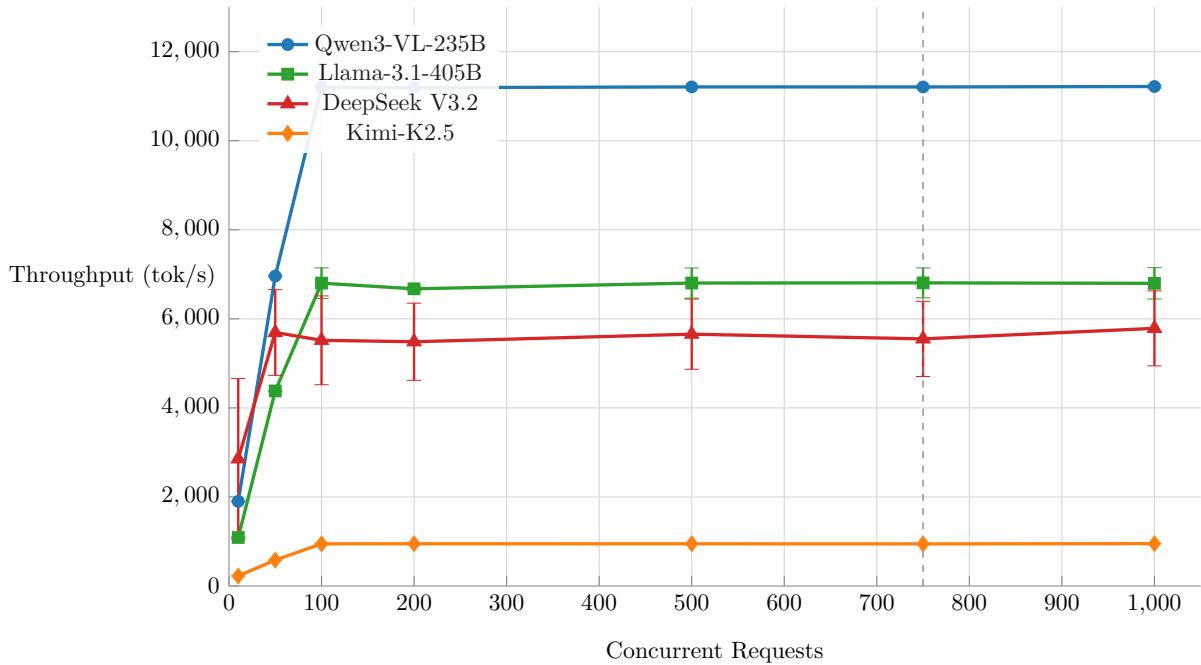
\begin{figure}[t]
\centering
\resizebox{\columnwidth}{!}{
\begin{tikzpicture}
\begin{axis}[
  width=\columnwidth,
  height=0.64\columnwidth,
  xlabel={Concurrent Requests},
  ylabel={Throughput (tok/s)},
  xmin=0, xmax=1050,
  ymin=0, ymax=13000,
  grid=major,
  grid style={gray!30, line width=0.5pt},
  legend pos=north west,
  legend style={font=\small, fill opacity=0.9, draw=none},
  every axis x label/.style={at={(0.5,-0.08)}, anchor=north},
  every axis y label/.style={at={(-0.12,0.5)}, anchor=south},
  scaled y ticks=false,
  yticklabel={\pgfmathprintnumber[fixed,1000 sep={,}]{\tick}},
  tick label style={font=\small},
  label style={font=\small},
  axis line style={gray!80},
  axis x line*=bottom,
  axis y line*=left,
]

\addplot[
  color={rgb,255:red,31;green,119;blue,180}, mark=*, mark size=2pt, line width=1.4pt,
  error bars/.cd, y dir=both, y explicit, error bar style={line width=1.0pt, color={rgb,255:red,31;green,119;blue,180}}, error mark options={rotate=90, mark size=3pt, color={rgb,255:red,31;green,119;blue,180}},
] coordinates {
  (10, 1902.3) +- (0, 6.7)
  (50, 6961.2) +- (0, 18.1)
  (100, 11198.3) +- (0, 13.7)
  (200, 11192.5) +- (0, 35.7)
  (500, 11208.7) +- (0, 9.5)
  (750, 11207.8) +- (0, 20.8)
  (1000, 11217.6) +- (0, 31.8)
};
\addlegendentry{Qwen3-VL-235B}

\addplot[
  color={rgb,255:red,44;green,160;blue,44}, mark=square*, mark size=2pt, line width=1.4pt,
  error bars/.cd, y dir=both, y explicit, error bar style={line width=1.0pt, color={rgb,255:red,44;green,160;blue,44}}, error mark options={rotate=90, mark size=3pt, color={rgb,255:red,44;green,160;blue,44}},
] coordinates {
  (10, 1090.1) +- (0, 8.8)
  (50, 4381.2) +- (0, 14.5)
  (100, 6802.0) +- (0, 343.2)
  (200, 6674.5) +- (0, 44.1)
  (500, 6804.3) +- (0, 338.4)
  (750, 6808.2) +- (0, 336.1)
  (1000, 6797.9) +- (0, 351.4)
};
\addlegendentry{Llama-3.1-405B}

\addplot[
  color={rgb,255:red,214;green,39;blue,40}, mark=triangle*, mark size=2.5pt, line width=1.4pt,
  error bars/.cd, y dir=both, y explicit, error bar style={line width=1.0pt, color={rgb,255:red,214;green,39;blue,40}}, error mark options={rotate=90, mark size=3pt, color={rgb,255:red,214;green,39;blue,40}},
] coordinates {
  (10, 2856.6) +- (0, 1803.0)
  (50, 5693.8) +- (0, 962.0)
  (100, 5518.2) +- (0, 997.4)
  (200, 5485.7) +- (0, 866.6)
  (500, 5656.5) +- (0, 788.4)
  (750, 5549.9) +- (0, 844.1)
  (1000, 5786.3) +- (0, 842.3)
};
\addlegendentry{DeepSeek V3.2}

\addplot[
  color={rgb,255:red,255;green,127;blue,14}, mark=diamond*, mark size=2.5pt, line width=1.4pt,
  error bars/.cd, y dir=both, y explicit, error bar style={line width=1.0pt, color={rgb,255:red,255;green,127;blue,14}}, error mark options={rotate=90, mark size=3pt, color={rgb,255:red,255;green,127;blue,14}},
] coordinates {
  (10, 225.1) +- (0, 1.3)
  (50, 582.9) +- (0, 2.7)
  (100, 947.6) +- (0, 5.6)
  (200, 949.6) +- (0, 7.2)
  (500, 948.2) +- (0, 4.6)
  (750, 946.5) +- (0, 6.4)
  (1000, 951.9) +- (0, 4.3)
};
\addlegendentry{Kimi-K2.5}

\draw[gray, dashed, line width=0.8pt, opacity=0.7] (axis cs:750,0) -- (axis cs:750,13000);

\end{axis}
\end{tikzpicture}}
\caption{Throughput vs.\ concurrency with 95\% confidence interval error bars from $n{=}5$ independent runs per model. DeepSeek~V3.2 exhibits the widest error bars (CoV up to 11.7\% at peak concurrency, reaching 50.8\% at concurrency 10), while Qwen3-VL and Kimi-K2.5 show near-deterministic behavior. Data from multi-run reproducibility workload (100 requests, 2{,}048 input / 512 output tokens); not directly comparable to the primary stress-test benchmark.}
\label{fig:reproducibility}
\end{figure}

Table~\ref{tab:reproducibility} reveals a substantial spread in measurement stability. Qwen3-VL exhibits near-deterministic behavior (CoV~$<$~0.3\% across all concurrency levels), indicating that throughput measurements for GQA-based MoE models are highly reproducible with minimal repetition. Llama-3.1-405B shows moderate variance (CoV~$\approx$~4\%), consistent with the larger memory footprint and denser compute patterns of a 405B-parameter model. DeepSeek~V3.2 exhibits the highest variance (CoV up to 11.7\% at peak concurrency, and reaching 50.8\% at concurrency~10 where individual run throughputs ranged from 1{,}485 to 5{,}193~tok/s), possibly attributable to MLA's latent-space projections and MoE routing stochasticity interacting with AITER kernel scheduling, though the specific mechanism has not been profiled. Kimi-K2.5 shows very low variance (CoV~$\approx$~0.4\%), comparable to Qwen3-VL despite also being an MLA-based MoE model; its stability may reflect the absence of AITER kernel scheduling non-determinism (AITER is disabled for Kimi on MI325X). These results suggest that single-run benchmarks are adequate for GQA models but that MLA-based models with AITER enabled benefit from multi-run averaging to obtain stable throughput estimates.

\subsection{Optimization Impact}
\label{subsec:optimization-impact}

Several vLLM and hardware-specific optimizations significantly influenced performance. We analyze their contributions below.

\subsubsection{AITER Kernel Acceleration}

The AMD AI Tensor Engine for ROCm (AITER) library provides optimized kernels for MoE and attention operations on CDNA~3 architecture. Table~\ref{tab:aiter-impact} summarizes AITER status across models.

\begin{table}[t]
\centering
\caption{AITER acceleration status across models. DeepSeek~V3.2 requires AITER for MLA attention on ROCm. Kimi-K2.5 cannot use AITER due to MLA head count incompatibility with MI325X.}
\label{tab:aiter-impact}
\begin{tabular}{@{}lllrl@{}}
\toprule
\textbf{Model} & \textbf{Workload} & \textbf{AITER} & \textbf{Peak tok/s} & \textbf{Notes} \\
\midrule
Llama-3.1-405B & Text   & Enabled  & 15,944 & MHA kernels (dense model) \\
DeepSeek V3.2  & Text   & Enabled  & 15,343 & MLA + AITER kernels \\
Qwen3-VL-235B  & Vision & Enabled  & 47,873 & MHA + MoE kernels \\
Kimi-K2.5      & Vision & Disabled & 7,327  & MLA head incompatibility \\
\bottomrule
\end{tabular}
\end{table}

DeepSeek~V3.2 with AITER enabled achieves higher throughput than Kimi-K2.5 without AITER; however, these models used different workloads (text vs.\ vision), making direct throughput comparison invalid. The throughput difference also reflects confounding factors including total parameter count (685B vs.\ 1T), quantization format (FP8 vs.\ INT4), tensor parallelism degree (TP$=$8 vs.\ TP$=$4), and GPU count.

\paragraph{MLA Ablation Infeasibility.}
We attempted a controlled ablation study comparing AITER-enabled and AITER-disabled inference for MLA models on MI325X. While a Triton MLA fallback exists (\texttt{VLLM\_ROCM\_USE\_AITER=0}), it delivers substantially lower performance than the AITER MLA backend, making it unsuitable as a controlled comparison (the performance gap conflates AITER's contribution with the fallback path's inherent limitations). Consequently, AITER's isolated contribution to MLA inference throughput cannot be cleanly measured on the current ROCm software stack.

\paragraph{GQA Model Ablation.}
Because MLA ablation is infeasible, we performed a controlled A/B comparison on Llama-3.1-405B (GQA attention, where AITER can be toggled). Both conditions use $n{=}5$ independent server restarts with an identical workload (100 requests, 2{,}048 input / 512 output tokens) at six concurrency levels. Table~\ref{tab:aiter-ablation} presents the comparison.

\begin{table}[t]
\centering
\caption{AITER ablation on Llama-3.1-405B (GQA). Both conditions use $n{=}5$ independent server restarts with identical workload. AITER provides a modest throughput benefit at single-request and high concurrency ($+$3--10\%), no benefit at mid-concurrency, and consistently higher variance.}
\label{tab:aiter-ablation}
\begin{tabular}{@{}rrrrrr@{}}
\toprule
\textbf{Conc.} & \textbf{Enabled tok/s} & \textbf{Disabled tok/s} & \textbf{Diff.} & \textbf{En.\ CoV} & \textbf{Dis.\ CoV} \\
\midrule
1   & 150 $\pm$ 9       & 137 $\pm$ 1       & $+$10.0\% & 4.69\% & 0.38\% \\
10  & 1{,}084 $\pm$ 23  & 1{,}092 $\pm$ 3   & $-$0.8\%  & 1.73\% & 0.23\% \\
50  & 4{,}340 $\pm$ 50  & 4{,}380 $\pm$ 12  & $-$0.9\%  & 0.93\% & 0.22\% \\
100 & 6{,}955 $\pm$ 136 & 6{,}682 $\pm$ 8   & $+$4.1\%  & 1.57\% & 0.10\% \\
200 & 6{,}871 $\pm$ 230 & 6{,}663 $\pm$ 17  & $+$3.1\%  & 2.69\% & 0.20\% \\
500 & 6{,}972 $\pm$ 137 & 6{,}676 $\pm$ 15  & $+$4.4\%  & 1.58\% & 0.18\% \\
\bottomrule
\end{tabular}
\end{table}

\begin{figure}[t]
\centering
\resizebox{\columnwidth}{!}{
\begin{tikzpicture}

\begin{axis}[
  name=top,
  width=\columnwidth,
  height=0.45\columnwidth,
  ybar,
  bar width=0.28cm,
  ylabel={Throughput (tok/s)},
  title={AITER Ablation: Llama-3.1-405B ($n{=}5$ runs per condition)},
  title style={font=\small},
  ymin=0, ymax=8500,
  xtick={0,1,2,3,4,5,6},
  xticklabels={},
  grid=major,
  grid style={gray!30, line width=0.5pt},
  ymajorgrids=true,
  xmajorgrids=false,
  scaled y ticks=false,
  yticklabel={\pgfmathprintnumber[fixed,1000 sep={,}]{\tick}},
  tick label style={font=\small},
  label style={font=\small},
  axis line style={gray!80},
  axis x line*=bottom,
  axis y line*=left,
  legend style={font=\small, at={(0.02,0.95)}, anchor=north west, draw=none, fill opacity=0.9},
  enlarge x limits=0.1,
]

\addplot[
  fill={rgb,255:red,44;green,160;blue,44}, draw=white, line width=0.8pt,
  error bars/.cd, y dir=both, y explicit, error bar style={line width=1.0pt}, error mark options={rotate=90, mark size=3pt},
] coordinates {
  (0, 150.2) +- (0, 8.1)
  (1, 548.7) +- (0, 3.9)
  (2, 1083.7) +- (0, 21.5)
  (3, 4340.2) +- (0, 46.5)
  (4, 6955.0) +- (0, 125.9)
  (5, 6871.1) +- (0, 212.8)
  (6, 6971.9) +- (0, 126.5)
};
\addlegendentry{AITER Enabled}

\addplot[
  fill={rgb,255:red,214;green,39;blue,40}, draw=white, line width=0.8pt,
  error bars/.cd, y dir=both, y explicit, error bar style={line width=1.0pt}, error mark options={rotate=90, mark size=3pt},
] coordinates {
  (0, 136.5) +- (0, 0.6)
  (1, 554.5) +- (0, 1.7)
  (2, 1092.1) +- (0, 2.8)
  (3, 4380.1) +- (0, 11.1)
  (4, 6682.1) +- (0, 7.7)
  (5, 6662.5) +- (0, 15.4)
  (6, 6676.0) +- (0, 13.5)
};
\addlegendentry{AITER Disabled}

\end{axis}

\begin{axis}[
  at={(top.below south west)},
  anchor=north west,
  yshift=-0.6cm,
  width=\columnwidth,
  height=0.26\columnwidth,
  ybar,
  bar width=0.45cm,
  ylabel={AITER Speedup (\%)},
  xlabel={Concurrency Level},
  xtick={0,1,2,3,4,5,6},
  xticklabels={1, 5, 10, 50, 100, 200, 500},
  xticklabel style={font=\small},
  ymin=-3, ymax=13,
  grid=major,
  grid style={gray!30, line width=0.5pt},
  ymajorgrids=true,
  xmajorgrids=false,
  tick label style={font=\small},
  label style={font=\small},
  axis line style={gray!80},
  axis x line*=bottom,
  axis y line*=left,
  enlarge x limits=0.1,
  clip=false,
  nodes near coords,
  nodes near coords align={above},
  every node near coord/.append style={font=\tiny\bfseries},
  point meta=explicit symbolic,
]

\draw[black, line width=0.6pt] (axis cs:-0.5,0) -- (axis cs:6.5,0);

\addplot[fill={rgb,255:red,44;green,160;blue,44}, draw=white, line width=0.8pt]
  coordinates {(0, 10.0) [{+10.0\%}]};
\addplot[fill={rgb,255:red,214;green,39;blue,40}, draw=white, line width=0.8pt, forget plot]
  coordinates {(1, -1.1) [{$-$1.1\%}]};
\addplot[fill={rgb,255:red,214;green,39;blue,40}, draw=white, line width=0.8pt, forget plot]
  coordinates {(2, -0.8) [{$-$0.8\%}]};
\addplot[fill={rgb,255:red,214;green,39;blue,40}, draw=white, line width=0.8pt, forget plot]
  coordinates {(3, -0.9) [{$-$0.9\%}]};
\addplot[fill={rgb,255:red,44;green,160;blue,44}, draw=white, line width=0.8pt, forget plot]
  coordinates {(4, 4.1) [{+4.1\%}]};
\addplot[fill={rgb,255:red,44;green,160;blue,44}, draw=white, line width=0.8pt, forget plot]
  coordinates {(5, 3.1) [{+3.1\%}]};
\addplot[fill={rgb,255:red,44;green,160;blue,44}, draw=white, line width=0.8pt, forget plot]
  coordinates {(6, 4.4) [{+4.4\%}]};

\end{axis}
\end{tikzpicture}}
\caption{AITER ablation on Llama-3.1-405B. Top: grouped bar chart comparing throughput with AITER enabled vs.\ disabled ($n{=}5$ runs, error bars show 95\% CI). Bottom: percentage speedup from AITER. AITER provides 10\% benefit at concurrency~1, negligible effect at mid-concurrency, and 3--5\% at high concurrency. Data from multi-run reproducibility workload (100 requests, 2{,}048 input / 512 output tokens).}
\label{fig:aiter-ablation}
\end{figure}
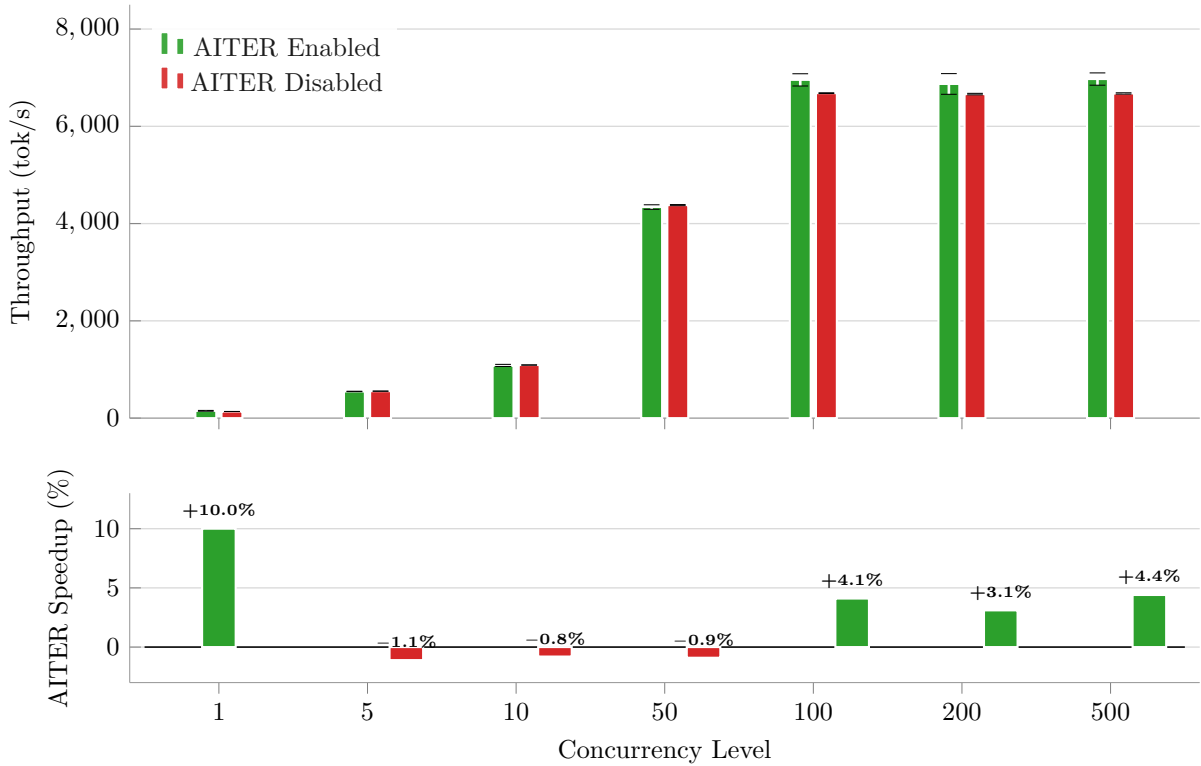

The ablation reveals a concurrency-dependent pattern. At single-request latency (concurrency~1), AITER provides a 10\% throughput improvement, likely from optimized single-stream kernel dispatch. At mid-concurrency (5--50), there is no meaningful difference ($<$1\%). At high concurrency (100--500), AITER provides a consistent 3--5\% throughput benefit. Across all concurrency levels, disabling AITER \emph{dramatically reduces measurement variability}: CoV drops from 0.6--4.7\% (enabled) to 0.1--0.4\% (disabled), a 2--16$\times$ reduction. This confirms that AITER kernel scheduling introduces non-determinism on CDNA~3 even for GQA models, consistent with the elevated variance observed for AITER-enabled DeepSeek~V3.2 (CoV~11.7\%, Table~\ref{tab:reproducibility}).

\paragraph{Implications.}
These results indicate that AITER's documented 2--3$\times$ speedups~\cite{amd2024aiter} are specific to MoE expert routing and MLA attention kernels; for standard GQA attention, the benefit is modest (3--10\% depending on concurrency). For GQA models, AITER's most pronounced effect is increased measurement variability. AITER's critical role for MLA models is as a \emph{performance enabler}, without which throughput drops to impractical levels via the Triton MLA fallback, making it a practical necessity for production deployments rather than merely an optional optimization. The AITER \texttt{VSKIP} optimization must be disabled on MI300X/MI325X (\texttt{AITER\_ENABLE\_VSKIP=0}) to prevent \texttt{HSA\_STATUS\_ERROR\_MEMORY\_APERTURE\_VIOLATION} errors in fused MoE kernels.

\subsubsection{Quantization and Memory Savings}

FP8 quantization provides approximately 50\% memory savings compared to FP16/BF16, enabling larger batch sizes and KV cache capacity. Table~\ref{tab:memory-precision} compares memory usage across precision formats.

\begin{table}[t]
\centering
\small
\caption{Per-GPU weight memory by precision format and load times on MI325X. Values from vLLM startup log. FP16/BF16 rows are theoretical estimates.}
\label{tab:memory-precision}
\begin{tabular}{@{}llrrl@{}}
\toprule
\textbf{Model} & \textbf{Precision} & \textbf{Per-GPU Memory} & \textbf{Load Time} & \textbf{Notes} \\
\midrule
DeepSeek V3.2  & FP8      & $\sim$83\,GiB  & $\sim$71s  & FP8 warmup +3\,min \\
DeepSeek V3.2  & FP16     & $\sim$180\,GiB & --         & 2.2$\times$ memory (est.) \\
Llama-3.1-405B & FP8      & $\sim$112\,GiB & $\sim$344s & Dense model \\
Llama-3.1-405B & BF16     & $\sim$224\,GiB & --         & 2$\times$ FP8 measured (est.) \\
Qwen3-VL-235B  & BF16     & $\sim$58\,GiB  & $\sim$112s & FP8 incompatible (ViT) \\
Kimi-K2.5      & INT4 QAT & $\sim$145\,GiB & $\sim$146s & Compressed-tensors \\
\bottomrule
\end{tabular}
\end{table}

FP8 quantization is particularly impactful for dense models: Llama-3.1-405B at BF16 would require an estimated ${\sim}$224\,GiB per GPU (extrapolating from the measured FP8 footprint of ${\sim}$112\,GiB), leaving minimal headroom within each GPU's 256\,GB capacity. FP8 halves the weight footprint, making single-node deployment feasible with substantial KV cache headroom. For MoE models, the savings are proportionally smaller because expert parameters are sparse, but FP8 still reduces DeepSeek~V3.2 from an estimated ${\sim}$180\,GiB to ${\sim}$83\,GiB per GPU.

The MI325X's 256\,GB HBM3e per GPU provides substantial headroom. DeepSeek~V3.2 in FP8 requires only ${\sim}$83\,GiB of per-GPU weight memory, consuming approximately 35\% of each GPU's capacity (noting that 83\,GiB $\approx$ 89\,GB) and leaving the remainder available for KV cache and batch buffers.

\paragraph{Memory Footprint Derivation.}
Table~\ref{tab:memory-derivation} presents a derivation of expected per-GPU weight memory from model parameters, quantization format, and tensor parallelism degree, compared against measured values from vLLM's startup log (``Model loading took X GiB'').\footnote{All ``Measured'' values are per-GPU weight memory as reported by vLLM's model loading log, confirmed via Phase~3 memory verification benchmarks. The ``Expected'' column computes $(\text{Stored Params} \times \text{Bytes per Param}) / \text{TP degree}$, which represents the naive sharding estimate. Discrepancies arise from embedding layer duplication across TP ranks, FP8 scale metadata, and communication buffer overhead.}

\begin{table}[t]
\centering
\small
\caption{Per-GPU weight memory derivation. ``Stored Params'' is the total count in GPU memory; for MoE models this equals all expert weights. Expected and Measured are per-GPU values (total divided by TP).}
\label{tab:memory-derivation}
\begin{tabular}{@{}lrrllrrr@{}}
\toprule
\textbf{Model} & \textbf{Total} & \textbf{Active} & \textbf{Stored} & \textbf{Precision} & \textbf{TP} & \textbf{Expected} & \textbf{Measured} \\
 & \textbf{Params} & \textbf{Params} & \textbf{Params} & & & \textbf{(GiB)} & \textbf{(GiB)} \\
\midrule
DeepSeek V3.2  & 685B & 37B  & 685B & FP8      & 8 & $\sim$87  & $\sim$83  \\
Llama-3.1-405B & 405B & 405B & 405B & FP8      & 8 & $\sim$52  & $\sim$112\textsuperscript{a} \\
Qwen3-VL-235B  & 235B & 22B  & 235B & BF16     & 8 & $\sim$59  & $\sim$58  \\
Kimi-K2.5      & 1T   & 32B  & 1T   & INT4 QAT & 4 & $\sim$140 & $\sim$145 \\
\bottomrule
\end{tabular}

\raggedright
\footnotesize
\textsuperscript{a}The 2.2$\times$ gap between expected (${\sim}$52\,GiB) and measured (${\sim}$112\,GiB) per GPU for Llama-3.1-405B is due to embedding layer duplication across TP ranks, FP8 scale factor metadata, and RCCL communication buffers.
\end{table}

The expected per-GPU memory footprint equals $(\text{Stored~Params} \times \text{Bytes/Param}) / \text{TP}$, where MoE models store all expert parameters even though only a subset are active per token. Effective bytes per parameter vary by format: FP8 requires ${\sim}1.02$~bytes (1~byte per weight plus scale factor overhead), BF16 requires 2.0~bytes, and INT4 QAT requires ${\sim}0.56$~bytes (0.5~bytes per weight plus FP16 scales per quantization group). Embedding layers and layer norms remain in BF16 regardless of quantization format but constitute $<$1\% of total parameters. In practice, measured per-GPU memory exceeds the naive estimate for dense models (Llama-3.1-405B) due to embedding table duplication across TP ranks and FP8 metadata; MoE models show closer agreement because expert weights dominate and shard evenly.

\subsubsection{KV Cache Management}

KV cache management strategies differ by attention mechanism:

\begin{itemize}
    \item \textbf{GQA models} (Qwen3-VL, Llama-3.1-405B): Support KV cache offloading to system RAM, enabling larger effective batch sizes. Qwen3-VL uses \texttt{--kv-offloading-backend native} with 64\,GB offloading, which contributes to its exceptional throughput. Llama-3.1-405B supports FP8 KV cache (\texttt{--kv-cache-dtype fp8}), though this was not enabled in our benchmarks.
    \item \textbf{MLA models} (DeepSeek~V3.2, Kimi-K2.5): Do not support KV cache offloading on the current ROCm stack (an MLA-specific limitation in vLLM~v0.14.1; vLLM's planned offloading redesign, RFC~\#22605, would also benefit MLA models). However, MLA inherently uses less KV cache memory per token, and the MI325X's large HBM capacity compensates.
\end{itemize}

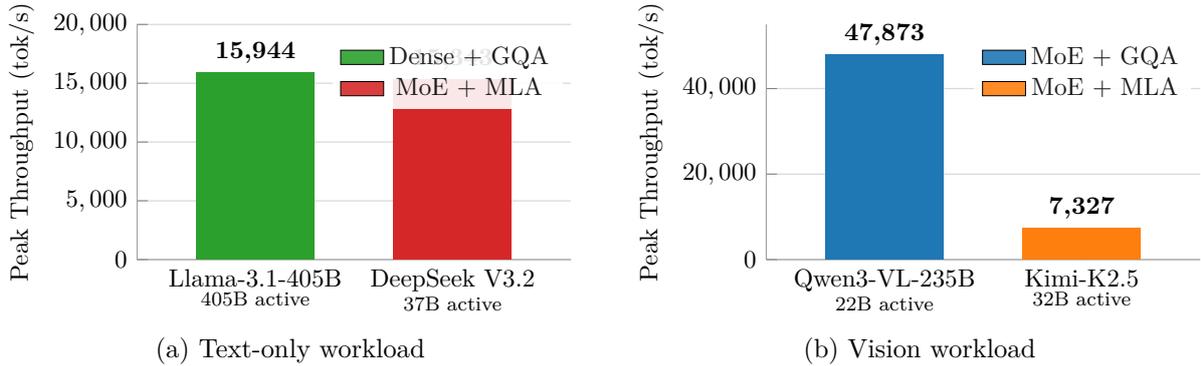
\begin{figure}[t]
\centering
\begin{minipage}[t]{0.48\columnwidth}
\centering
\resizebox{\textwidth}{!}{
\begin{tikzpicture}
\begin{axis}[
  width=\columnwidth,
  height=0.64\columnwidth,
  ylabel={Peak Throughput (tok/s)},
  ymin=0, ymax=20000,
  xmin=-0.6, xmax=1.6,
  xtick={0,1},
  xticklabels={%
    \shortstack{Llama-3.1-405B\\[-1pt]{\scriptsize 405B active}},
    \shortstack{DeepSeek V3.2\\[-1pt]{\scriptsize 37B active}}%
  },
  xticklabel style={font=\small, align=center},
  grid=major,
  grid style={gray!30, line width=0.5pt},
  ymajorgrids=true,
  xmajorgrids=false,
  scaled y ticks=false,
  yticklabel={\pgfmathprintnumber[fixed,1000 sep={,}]{\tick}},
  tick label style={font=\small},
  label style={font=\small},
  axis line style={gray!80},
  axis x line*=bottom,
  axis y line*=left,
  legend style={font=\small, at={(0.98,0.95)}, anchor=north east, draw=none, fill opacity=0.9},
  clip=false,
]

\fill[fill={rgb,255:red,44;green,160;blue,44}]
  (axis cs:-0.3,0) rectangle (axis cs:0.3,15944);
\node[above, font=\small\bfseries] at (axis cs:0,15944) {15,944};

\fill[fill={rgb,255:red,214;green,39;blue,40}]
  (axis cs:0.7,0) rectangle (axis cs:1.3,15343);
\node[above, font=\small\bfseries] at (axis cs:1,15343) {15,343};

\addlegendimage{area legend, fill={rgb,255:red,44;green,160;blue,44}}
\addlegendentry{Dense + GQA}
\addlegendimage{area legend, fill={rgb,255:red,214;green,39;blue,40}}
\addlegendentry{MoE + MLA}

\end{axis}
\end{tikzpicture}}
\subcaption{Text-only workload}
\end{minipage}
\hfill
\begin{minipage}[t]{0.48\columnwidth}
\centering
\resizebox{\textwidth}{!}{
\begin{tikzpicture}
\begin{axis}[
  width=\columnwidth,
  height=0.64\columnwidth,
  ylabel={Peak Throughput (tok/s)},
  ymin=0, ymax=55000,
  xmin=-0.6, xmax=1.6,
  xtick={0,1},
  xticklabels={%
    \shortstack{Qwen3-VL-235B\\[-1pt]{\scriptsize 22B active}},
    \shortstack{Kimi-K2.5\\[-1pt]{\scriptsize 32B active}}%
  },
  xticklabel style={font=\small, align=center},
  grid=major,
  grid style={gray!30, line width=0.5pt},
  ymajorgrids=true,
  xmajorgrids=false,
  scaled y ticks=false,
  yticklabel={\pgfmathprintnumber[fixed,1000 sep={,}]{\tick}},
  tick label style={font=\small},
  label style={font=\small},
  axis line style={gray!80},
  axis x line*=bottom,
  axis y line*=left,
  legend style={font=\small, at={(0.98,0.95)}, anchor=north east, draw=none, fill opacity=0.9},
  clip=false,
]

\fill[fill={rgb,255:red,31;green,119;blue,180}]
  (axis cs:-0.3,0) rectangle (axis cs:0.3,47873);
\node[above, font=\small\bfseries] at (axis cs:0,47873) {47,873};

\fill[fill={rgb,255:red,255;green,127;blue,14}]
  (axis cs:0.7,0) rectangle (axis cs:1.3,7327);
\node[above, font=\small\bfseries] at (axis cs:1,7327) {7,327};

\addlegendimage{area legend, fill={rgb,255:red,31;green,119;blue,180}}
\addlegendentry{MoE + GQA}
\addlegendimage{area legend, fill={rgb,255:red,255;green,127;blue,14}}
\addlegendentry{MoE + MLA}

\end{axis}
\end{tikzpicture}}
\subcaption{Vision workload}
\end{minipage}
\caption{Peak throughput comparison separated by workload type. (a)~Text-only workload (500-token input, 100-token output). (b)~Vision workload (100-token text + 1 image, 200-token output). Total tok/s for vision models includes image tokens processed by the vision encoder and is not directly comparable to text models. Data from primary stress-test benchmark ($3\times$ multiplier).}
\label{fig:peak-throughput-bar}
\end{figure}

\paragraph{Key Insight.} Within each workload type, active parameter count per token is consistently associated with inference throughput. In the text workload, DeepSeek~V3.2 (37B active) matches Llama-3.1-405B (405B active) in total throughput despite using only 9\% of the active parameters. In the vision workload, Qwen3-VL (22B active) achieves 6.5$\times$ the throughput of Kimi-K2.5 (32B active). These patterns are consistent with LLM inference on MI325X being fundamentally memory-bandwidth-bound: fewer active parameters means less data movement per token. However, these comparisons are confounded by differences in quantization format, AITER acceleration status, and tensor parallelism degree, so the active-parameter relationship cannot be fully isolated.

\section{Discussion}
\label{sec:discussion}

Our benchmark results reveal several findings with implications for production LLM deployment on AMD hardware and for the broader understanding of inference serving across diverse model architectures.

\subsection{Architecture-Aware Optimization: One Size Does Not Fit All}

A central finding is that serving configuration must be tailored to model architecture; a single default configuration cannot achieve optimal (or even correct) results across the architectures we evaluate. The divergence between MLA and GQA models is particularly stark:

\begin{itemize}
    \item \textbf{MLA constraints.} Both MLA models (DeepSeek~V3.2 and Kimi-K2.5) require \texttt{--block\allowbreak{}-size~1} on the ROCm/AITER stack for their compressed latent KV cache (other platforms use larger block sizes), are incompatible with KV cache offloading on the current ROCm stack (vLLM's offloading redesign, RFC~\#22605, would also benefit MLA models), and impose specific attention head distribution requirements for tensor parallelism. Kimi-K2.5 is further constrained to TP$=$4 in our evaluation (see Section~\ref{subsec:tensor-parallelism} for the rationale), forgoing half the cluster's bandwidth and memory.
    \item \textbf{GQA flexibility.} GQA models (Llama-3.1-405B and Qwen3-VL-235B) benefit from KV cache offloading, support FP8 KV cache quantization (where vision encoder constraints permit), and operate with standard block sizes. This flexibility contributes to Qwen3-VL's 47{,}873~tok/s peak under its vision workload. Among vision models, Qwen3-VL achieves 6.5$\times$ the throughput of Kimi-K2.5, while among text models, Llama-3.1-405B (also GQA) matches DeepSeek~V3.2 (MLA) at comparable throughput despite 10$\times$ more active parameters.
    \item \textbf{AITER selectivity.} AITER is required for competitive production MLA inference throughput on ROCm; a Triton MLA fallback exists but delivers substantially lower performance, making controlled ablation impractical (the performance gap conflates AITER's contribution with the fallback path's limitations). Our controlled ablation on Llama-3.1-405B (GQA, $n{=}5$ per condition) demonstrates that AITER provides a modest throughput benefit for standard attention models: 3--5\% at high concurrency and ${\sim}$10\% in total throughput at single-request latency (though output throughput, i.e.\ pure generation speed, shows no meaningful difference, indicating the gain is confined to prefill acceleration), but no benefit at mid-concurrency (Table~\ref{tab:aiter-ablation}), while increasing measurement variability by 2--16$\times$ (coefficient of variation) (CoV 0.6--4.7\% enabled vs.\ 0.1--0.4\% disabled). This confirms that AMD's documented 2--3$\times$ speedups~\cite{amd2024aiter} are specific to MoE and MLA kernels; GQA models see only single-digit percentage gains. AITER must be entirely disabled for Kimi-K2.5 due to MXFP4 hardware requirements (CDNA~4 only) and head count incompatibilities. Even for AITER-compatible deployments, \texttt{VSKIP} must be disabled (a known issue documented in AITER Issue~\#1143) to prevent \texttt{HSA\_STATUS\_ERROR\_MEMORY\_APERTURE\_VIOLATION} errors in fused MoE kernels on CDNA~3.
\end{itemize}

These findings challenge the implicit assumption in many deployment guides that a single set of vLLM flags suffices for all models. Production deployments should implement architecture-detection logic that selects block size, AITER configuration, KV cache strategy, and tensor parallelism degree based on the model's attention mechanism and expert structure.

\subsection{AMD MI325X Platform Viability}

The MI325X cluster demonstrates strong viability as an inference platform for frontier-scale models. Several hardware characteristics prove particularly advantageous:

\paragraph{Memory capacity eliminates offloading necessity.} The 256\,GB HBM3e per GPU (2\,TB aggregate) is sufficient to host all models tested, including the 1T-parameter Kimi-K2.5, with substantial headroom for KV cache. DeepSeek~V3.2 in FP8 requires only ${\sim}$83\,GiB of per-GPU weight memory (as reported by vLLM's model loading log), consuming approximately 35\% of each GPU's 256\,GB capacity (noting that 83\,GiB $\approx$ 89\,GB) and leaving the remainder for KV cache and batch state. This eliminates the need for CPU memory offloading in most practical scenarios, simplifying deployment architecture and removing offloading-induced latency.

\paragraph{Aggregate bandwidth supports high throughput.} The 48\,TB/s aggregate memory bandwidth (6.0\,TB/s per GPU) enables the high token throughput observed across all models. Qwen3-VL's 47{,}873 tok/s peak (vision workload, including $\sim$1{,}000 image tokens per request) demonstrates that the platform can sustain high throughput for frontier-scale multimodal models. For text-only workloads, the platform achieves up to 15,944 tok/s (Llama-3.1-405B).

\paragraph{Trillion-parameter model support.} To our knowledge, this study represents the first published inference benchmark of a trillion-parameter model on MI325X (CDNA~3). AMD has published Kimi-K2 benchmarks on MI355X (CDNA~4), and Oak Ridge National Lab trained a trillion-parameter model on MI250X GPUs~\cite{arXiv_2312_12705}. Kimi-K2.5's~\cite{kimik2_2025} 1T parameters fit within four MI325X GPUs using INT4 QAT quantization, achieving 7,327 tok/s at 500 concurrent requests with 100\% reliability through 1,000 concurrent. This demonstrates that the MI325X platform can serve the largest publicly available models without multi-node communication overhead.

\paragraph{Power and thermal behavior under load.} GPU-level monitoring during the Kimi-K2.5 benchmark runs (3 runs, 1-second sampling via \texttt{rocm-smi}) provides insight into power and thermal characteristics. Table~\ref{tab:power-thermal} summarizes the per-GPU power and thermal measurements.

\begin{table}[t]
\centering
\caption{Power and thermal characteristics during Kimi-K2.5 inference (TP$=$4, 3 runs, 1-second sampling via \texttt{rocm-smi}). Active GPUs are the 4 GPUs participating in tensor parallelism; idle GPUs are the remaining 4 GPUs on the node.}
\label{tab:power-thermal}
\begin{tabular}{@{}l rr@{}}
\toprule
\textbf{Metric} & \textbf{Active GPUs (4)} & \textbf{Idle GPUs (4)} \\
\midrule
Mean power (W/GPU) & 662 & 118 \\
Min power (W/GPU) & 129 & --- \\
Max power (W/GPU) & 798 & --- \\
\% of TDP (1{,}000\,W) & ${\sim}$66\% & ${\sim}$12\% \\
\midrule
Mean junction temp (\textdegree{}C) & 64 & --- \\
Peak junction temp (\textdegree{}C) & 76 & --- \\
Mean HBM3e temp (\textdegree{}C) & 56 & --- \\
Peak HBM3e temp (\textdegree{}C) & 66 & --- \\
\midrule
Hardware utilization & ${\sim}$99.5\% & --- \\
FLOPs utilization & 1.1\% & --- \\
\bottomrule
\end{tabular}
\end{table}

The four idle GPUs consumed a combined 472\,W, representing ${\sim}$15\% of total system GPU power (3.1\,kW) for zero productive work. This quantifies the practical cost of MLA's tensor parallelism constraint: the TP$=$4 requirement leaves half the node's GPUs idle and their associated power budget unutilized. Both junction and HBM3e temperatures remain well within the MI325X's thermal envelope, indicating substantial thermal headroom under production workloads. Active GPUs reported ${\sim}$99.5\% hardware utilization despite achieving only 1.1\% FLOPs utilization (Table~\ref{tab:flops-utilization}), reinforcing that the GPUs are fully occupied servicing memory transfers rather than compute operations, a direct consequence of the memory-bandwidth-bound nature of autoregressive inference discussed in Section~\ref{sec:results}.

\subsection{Memory Bandwidth as the Primary Bottleneck}

Under the primary stress test workloads (500-token input, 100-token output for text models; 100-token input + 1 image, 200-token output for vision models), all four models exhibit throughput saturation at approximately 500 concurrent requests on our 8-GPU MI325X cluster, despite spanning a 4$\times$ range in total parameters and employing three different architectural paradigms. However, this saturation point is workload-dependent: under the multi-run workload (2{,}048-token input, 512-token output), saturation occurs earlier, at approximately 100--200 concurrent requests, as longer sequences consume more memory bandwidth per request. The consistent saturation across architecturally diverse models within a given workload, combined with the sub-15\% FLOPs utilization (Table~\ref{tab:flops-utilization}), strongly suggests that the bottleneck at high concurrency is memory bandwidth rather than compute capacity. The MI325X cluster provides 10.5 PFLOPS of FP16 compute but only 48\,TB/s of aggregate memory bandwidth. Recasens et al.~\cite{arXiv_2503_08311} have shown that DRAM bandwidth saturation dynamics drive model-dependent saturation on other hardware platforms, and the LIMINAL analytical model~\cite{davies2025liminal} identifies memory bandwidth as the primary barrier for LLM decode performance across GPU architectures. Table~\ref{tab:flops-utilization} quantifies the compute utilization for each model at peak output throughput.

\begin{table}[t]
\centering
\caption{FLOPs utilization at peak output throughput. Each output token requires approximately $2 \times P_{\text{active}}$ FLOPs. MI325X peak compute: 2,615\,TFLOPS FP8 and 1,307\,TFLOPS BF16 per GPU.}
\label{tab:flops-utilization}
\begin{tabular}{@{}llrrrr@{}}
\toprule
\textbf{Model} & \textbf{Active} & \textbf{Precision} & \textbf{GPUs} & \textbf{Output tok/s} & \textbf{Utilization} \\
\midrule
Qwen3-VL-235B  & 22B  & BF16    & 8 & 7,140 & 3.0\%  \\
Llama-3.1-405B & 405B & FP8     & 8 & 3,673 & 14.2\% \\
DeepSeek V3.2  & 37B  & FP8     & 8 & 1,239 & 0.4\%  \\
Kimi-K2.5      & 32B  & INT4 QAT\textsuperscript{$\dagger$} & 4 & 867   & 1.1\%  \\
\bottomrule
\end{tabular}

\raggedright
\footnotesize
\textsuperscript{$\dagger$}INT4 QAT weights are dequantized to BF16 for compute; utilization computed against BF16 peak (1,307\,TFLOPS/GPU).
\end{table}

Even the most compute-intensive model (Llama-3.1-405B) utilizes only 14\% of peak FP8 compute, while MoE models operate at 0.4--3\% utilization, confirming that autoregressive decoding is overwhelmingly memory-bandwidth-bound~\cite{pope2023efficiently}.

This has several implications. First, for a given workload, the common saturation point means that adding more concurrent requests beyond the saturation threshold yields no throughput benefit. For our stress test workload, this threshold is ${\sim}$500; for longer-sequence workloads, saturation occurs earlier. Second, optimizations that reduce memory traffic, such as MoE sparsity (fewer active parameters per token), quantization (fewer bytes per parameter), and MLA's compressed KV cache (smaller memory reads per attention operation), are more impactful than optimizations that increase raw compute throughput. Third, the 100\% success rate at saturation (no request failures at 1,000 concurrent) indicates that vLLM's scheduling and queuing mechanisms~\cite{kwon2023vllm} gracefully handle overload conditions, converting excess concurrency into increased latency rather than failures.

\subsection{Transferability to AMD Instinct MI300X}

Our findings transfer directly to the AMD Instinct MI300X because both accelerators share the CDNA~3 microarchitecture (gfx942 ISA) and identical compute unit count (304 CUs). All software-level configurations (AITER kernel flags, block size constraints, tensor parallelism requirements) apply without modification. The primary hardware differences are memory capacity (192\,GB HBM3 on MI300X vs.\ 256\,GB HBM3e on MI325X) and memory bandwidth (5.3\,TB/s vs.\ 6.0\,TB/s). All four models fit on an 8-GPU MI300X node. However, the MI300X's ${\sim}$12\% lower bandwidth should translate to a roughly proportional reduction in peak throughput, and reduced per-GPU capacity leaves less headroom for KV cache at high concurrency.

\subsection{Limitations}
\label{subsec:limitations}

This study has several limitations that should be considered when interpreting the results:

\paragraph{Single cluster configuration.} All experiments were conducted on a single 8-GPU MI325X node. Multi-node scaling behavior, which introduces inter-node communication overhead and pipeline parallelism considerations, remains unexplored. The TP$=$4 constraint for Kimi-K2.5 further limits our evaluation to half the cluster for that model.

\paragraph{Model selection.} We evaluate four models across three architectural families, but many important architectures remain untested, including standard transformer models smaller than 100B parameters, state-space models (Mamba~\cite{gu2023mamba}), and hybrid architectures. Our findings about architecture-specific optimization may not generalize to all model families.

\paragraph{Single-platform evaluation.} This study focuses exclusively on the AMD MI325X platform. Cross-platform comparisons require identical workloads, software versions, and measurement methodology, and are beyond the scope of this work.

\paragraph{Single vLLM version.} All experiments use vLLM~v0.14.1 (with nightly builds for Kimi-K2.5). The vLLM project evolves rapidly, and performance characteristics may change significantly across versions as ROCm support matures, AITER kernels are updated, and scheduling algorithms improve. Our architecture-specific configuration findings (block size, AITER flags) are version-dependent and should be re-validated on future releases.

\paragraph{Evaluation scope.} This study evaluates serving throughput and tail latency under concurrent load. Time to First Token (TTFT) and output quality metrics are orthogonal concerns studied elsewhere~\cite{deepseekv3_2024, kimik2_2025} and are outside the scope of this work.

\paragraph{Software maturity considerations.} The ROCm and AITER ecosystem is production-capable and continues to mature, with active development in areas that will further streamline deployment. Remaining considerations include the \texttt{VSKIP} workaround requiring explicit disabling, vision encoder FP8 incompatibilities due to dimension constraints, and the need for nightly builds for Kimi-K2.5 support. These are active areas of development expected to be addressed in subsequent ROCm and AITER releases.

\subsection{Practical Deployment Considerations}

Our experience deploying and benchmarking these models yields several practical recommendations for production MI325X deployments:

\begin{enumerate}
    \item \textbf{Architecture detection before configuration.} Determine the model's attention mechanism (MLA vs.\ GQA), sparsity (dense vs.\ MoE), and expert count before selecting vLLM flags. On the current ROCm stack, MLA models require \texttt{--block\allowbreak{}-size~1} and cannot use \texttt{--kv-offloading-\allowbreak{}backend~native}.
    \item \textbf{AITER validation per model.} Enable AITER globally (\texttt{VLLM\_ROCM\_USE\_AITER=1}) but validate correctness for each model. Set \texttt{AITER\_ENABLE\_VSKIP=0} for all MLA deployments. Disable AITER entirely for models with incompatible attention head counts.
    \item \textbf{Quantization format compatibility.} Verify FP8 kernel compatibility before assuming FP8 quantization is available. Vision encoder dimension constraints and MLA-specific KV cache formats may force alternative precision formats.
    \item \textbf{Concurrency planning.} For short-sequence workloads (${\sim}$500-token input), target approximately 500 concurrent requests for maximum throughput on the 8-GPU MI325X cluster; saturation occurs earlier (${\sim}$100--200 concurrent) for longer sequences (2{,}048+ tokens). Additional concurrency beyond the saturation point adds latency without throughput benefit.
    \item \textbf{Container and build management.} Maintain separate container images for stable and nightly vLLM builds, as some frontier models require nightly support before stable release integration.
\end{enumerate}

\section{Conclusion and Future Work}
\label{sec:conclusion}

We have presented a systematic cross-architecture benchmark study of large language model inference on AMD Instinct MI325X GPUs, evaluating four frontier-scale models spanning 235 billion to 1 trillion parameters on an 8-GPU cluster with 2\,TB aggregate HBM3e using the vLLM serving framework.

\subsection{Summary of Contributions}

Our study makes five key contributions:

\begin{enumerate}
    \item \textbf{Cross-architecture MI325X inference benchmark at scale.} We provide systematic performance characterization for four models across three architectural families (dense GQA, MoE+GQA, MoE+MLA) under workloads from single-request latency to 1{,}000 concurrent users, processing 18.9~million tokens across 17{,}406 requests with 100\% HTTP-level success.

    \item \textbf{Architecture-aware optimization is essential.} We demonstrate that MLA and GQA architectures require fundamentally different serving configurations. On the current ROCm stack, MLA models require block size~1 and cannot use KV cache offloading; GQA models benefit from offloading and standard block sizes. The AITER runtime is required for competitive production MLA inference throughput on ROCm; a Triton MLA fallback exists but delivers substantially lower performance, making AITER a practical necessity for production deployments. A controlled ablation on Llama-3.1-405B (GQA, $n{=}5$ per condition) shows a modest 3--5\% throughput benefit at high concurrency but 2--16$\times$ higher measurement variability (coefficient of variation) with AITER enabled (Table~\ref{tab:aiter-ablation}), confirming that AITER's large speedups target MoE and MLA kernels specifically. AITER must be entirely disabled for Kimi-K2.5 due to MXFP4 hardware requirements (CDNA~4 only) and attention head count constraints.

    \item \textbf{Trillion-parameter model on MI325X (CDNA~3).} We deploy and benchmark Kimi-K2.5~\cite{kimik2_2025}, a 1T-parameter MoE model, on four MI325X GPUs using INT4 QAT quantization, achieving 7,327 tok/s at 500 concurrent requests with 100\% reliability through 1,000 concurrent.

    \item \textbf{Active parameter count is associated with throughput.} Active parameter count per token is more consistently associated with inference throughput than total parameter count within each workload type, though confounded by other experimental variables (Section~\ref{sec:results}). Within the text workload, DeepSeek~V3.2 (MoE+MLA, 37B active) matches Llama-3.1-405B (Dense+GQA, 405B active) at 15,343 vs.\ 15,944 tok/s despite having only 9\% of the active parameters. Within the vision workload, Qwen3-VL-235B (MoE+GQA, 22B active) achieves 6.5$\times$ the throughput of Kimi-K2.5 (MoE+MLA, 32B active).

    \item \textbf{Workload-dependent throughput saturation.} All four models, despite a 4$\times$ range in total parameters and three distinct architectural paradigms, exhibit a common throughput saturation point within a given workload on our 8-GPU MI325X cluster, consistent with a memory-bandwidth bottleneck. The saturation threshold is workload-dependent: ${\sim}$500 concurrent for the stress test workload (500-token input, 100-token output for text; 100-token input + 1 image, 200-token output for vision) and ${\sim}$100--200 concurrent for longer-sequence workloads (2{,}048-token input, 512-token output). Prior work has shown that DRAM bandwidth saturation dynamics drive model-dependent saturation on other hardware platforms~\cite{arXiv_2503_08311}.
\end{enumerate}

\subsection{Future Work}

Several directions would extend and strengthen the findings presented here:

\paragraph{Multi-node scaling.} Our evaluation is limited to a single 8-GPU node. Extending to multi-node configurations would introduce inter-node communication overhead (InfiniBand or RoCE) and enable evaluation of pipeline parallelism and expert parallelism strategies for MoE models. Understanding whether the ${\sim}$500-concurrent saturation point shifts proportionally with additional nodes is of direct practical interest.

\paragraph{Pipeline and expert parallelism.} The current study uses tensor parallelism exclusively. Pipeline parallelism could enable higher throughput for dense models like Llama-3.1-405B by overlapping computation across pipeline stages, while expert parallelism could improve MoE model serving by distributing experts across GPUs independently of the attention parallelism degree.

\paragraph{Broader model coverage.} Extending the evaluation to additional architectures, including state-space models (e.g., Mamba~\cite{gu2023mamba}), hybrid attention-SSM architectures, and smaller models in the 7B--70B range, would test the generality of our architecture-aware optimization findings and active-parameter throughput relationship.

\paragraph{Speculative decoding.} Speculative decoding~\cite{leviathan2023speculativedecoding} is orthogonal to our evaluated optimizations and could further improve single-request latency and throughput, particularly for MLA models where per-token generation speed is lower. Evaluating draft-model selection and acceptance rates across our four architectures would complement the throughput-focused results presented here.

\paragraph{Disaggregated prefill and decode.} Separating prefill and decode phases onto different GPU pools (an emerging approach in production serving systems) could improve overall throughput by dedicating compute-optimized GPUs to prefill and bandwidth-optimized GPUs to decode. The MI325X's high memory bandwidth would make it a strong candidate for the decode phase, suggesting a heterogeneous deployment where prefill runs on compute-dense accelerators.

\paragraph{Streaming inference and per-token latency characterization.} Our benchmarks use non-streaming requests, measuring end-to-end latency and aggregate throughput but not per-token metrics such as Time to First Token (TTFT) and Inter-Token Latency (ITL). These metrics are critical for interactive applications where perceived responsiveness depends on how quickly the first token arrives and how smoothly subsequent tokens stream. Re-running the benchmark suite with streaming enabled would characterize prefill latency (TTFT) and decode cadence (ITL) across all four architectures, potentially revealing architecture-specific differences; for example, whether MLA's compressed KV cache format affects prefill latency differently than GQA's standard attention.

\paragraph{Comprehensive power and thermal profiling.} GPU-level power and thermal monitoring in this study was collected only for Kimi-K2.5 (Table~\ref{tab:power-thermal}). Extending \texttt{rocm-smi} monitoring to all four models across the full concurrency range would enable cross-architecture power efficiency comparisons (tokens per watt), characterize how power draw scales with concurrency, and quantify the energy cost differences between dense, MoE+GQA, and MoE+MLA inference. This data would also inform capacity planning for datacenter power budgets.

\paragraph{Automated architecture-aware configuration.} Our findings on architecture-specific optimization requirements motivate the development of automated configuration selection systems that detect model architecture (attention mechanism, expert structure, head count) and automatically select optimal vLLM parameters, AITER flags, tensor parallelism degree, and quantization format. Such a system would reduce the deployment expertise barrier and prevent the silent correctness and performance issues we encountered during manual configuration.

\section*{Acknowledgments}
The authors thank Vultr for providing the 8-GPU AMD Instinct MI325X cluster used in this study. We also thank the open-source vLLM and ROCm communities for their contributions to the inference software stack.

\bibliographystyle{plainnat}
\bibliography{references}

\end{document}